\documentclass[sigconf, nonacm]{acmart}

\usepackage{amssymb}
\usepackage{amsmath,amsfonts}
\usepackage{graphicx}
\usepackage{textcomp}
\usepackage{url}
\usepackage{multirow}
\usepackage{fancyhdr}
\usepackage{dblfloatfix}
\usepackage{makecell}
\usepackage{xspace}
\usepackage{enumitem}
\PassOptionsToPackage{bookmarks=true,breaklinks=true,letterpaper=true,colorlinks,citecolor=blue,linkcolor=blue,urlcolor=blue}{hyperref}
\usepackage{hyperref}
\newcommand{\namePaper}{{Harmonia}} %

\newcommand{\sibyl}{{Sibyl}}
\newcommand{\cde}{{CDE}}
\newcommand{\ksvm}{{K-SVM}}
\newcommand{\rnn}{{RNN-HSS}}
\newcommand{\sapm}{{SAPM}}
\newcommand{\oracle}{{Oracle}}
\newcommand{\rlmigr}{{RL-Migr}}

\newcommand{\figs}[1]{{Figs.~#1}} %
\newcommand{\fig}[1]{{Fig.~#1}} %
\newcommand{\tab}[1]{{Table~#1}} %

\newcommand{\head}[1]{{\noindent\textbf{#1.}}} %

\newcommand*\circled[1]{%
  \begin{picture}(8,8)(1,2)%
    \put(5,5){\color{black}\circle*{8}}%
    \put(5,5){\makebox(0,0){\color{white}\bfseries\small#1}}%
  \end{picture}%
}

\newcommand\scalemath[2]{\scalebox{#1}{\mbox{\ensuremath{\displaystyle #2}}}}

\newcommand{\etal}{\textit{et al.}}
\newif\ifsubmission
\submissiontrue

\let\marginpar\marginnote
\newcommand{\cami}[1]{{\color{black}#1}}
\newcommand{\camii}[1]{{\color{black}#1}}
\newcommand{\camiii}[1]{{\color{black}#1}}
\newcommand{\camiv}[1]{{\color{black}#1}}

\AtBeginDocument{%
  }

\title[\namePaper{}: Enhancing Data Placement and Migration in HSS via Multi-Agent RL]{\namePaper{}: Enhancing Data Placement and Migration \\
in Hybrid Storage Systems via Multi-Agent \\
Reinforcement Learning}

\newcommand{\affilETH}[0]{\textsuperscript{\S}}
\newcommand{\affilAMD}[0]{\textsuperscript{$\dagger$}}
\newcommand{\affilPostec}[0]{\textsuperscript{$\nabla$}}

\author{Rakesh Nadig\affilETH \hspace{0.5cm} Vamanan Arulchelvan\affilETH \hspace{0.5cm}  Rahul Bera\affilETH \hspace{0.5cm}  
Taha Shahroodi \affilETH \hspace{0.5cm}  Gagandeep Singh\affilAMD \hspace{0.5cm} \\ Andreas Kakolyris \affilETH \hspace{0.5cm} İsmail Emir Yüksel \affilETH \hspace{0.5cm} Mohammad Sadrosadati \affilETH \hspace{0.5cm} Jisung Park\affilPostec \hspace{0.5cm} \\ Onur Mutlu\affilETH \\ \vspace{0.2cm}
\emph{\affilETH ETH Z{\"u}rich \hspace{1cm} \affilAMD AMD Research \hspace{1cm} \affilPostec POSTECH \hspace{1cm}\\}
}

\renewcommand{\shortauthors}{}

\renewcommand{\authors}{Rakesh Nadig, Vamanan Arulchelvan, Rahul Bera, Taha Shahroodi, Gagandeep Singh, Andreas Kakolyris, İsmail Emir Yüksel, Mohammad Sadrosadati, Jisung Park, Onur Mutlu}

\fancypagestyle{firstpage}{
    \fancyhf{}
    \fancyfoot[C]{\large\thepage}
}

\begin{document}

\begin{abstract}
Modern high-performance computing (HPC) environments rely on hybrid storage systems (HSS) that combine multiple storage devices with diverse latency, bandwidth, endurance, and capacity characteristics to meet the performance, capacity, and cost requirements of data-intensive applications. 
The performance of an HSS highly depends on two key data-management policies: (1) \emph{data placement}, which determines the most suitable storage device to store application data, and (2) \emph{data migration}, which dynamically reorganizes previously-stored data across storage devices (i.e., prefetching hot data and evicting cold data) to sustain high HSS performance.
These policies are tightly interdependent, and thus, improving one without considering the other leads to suboptimal HSS performance.
Unfortunately, prior works focus on optimizing \emph{only} one of the policies.  

Our goal is to design a holistic data-management technique that optimizes both data-placement and data-migration policies to fully exploit the potential of an HSS. 
To this end, we propose \textbf{\namePaper{}}, a multi-agent reinforcement learning (RL)-based data-management technique. 
\namePaper{} employs two lightweight autonomous RL agents, a data-placement agent and a data-migration agent, that adapt their policies for the current workload and HSS configuration while coordinating with each other. 

We evaluate \namePaper{} on real HSS configurations with up to four heterogeneous storage devices and 25 data-intensive workloads.
On \cami{a} performance- (cost-) optimized HSS with two heterogeneous storage devices, \namePaper{} outperforms the best-performing prior approach by 29.3\% (44.8\%) on average.
On an HSS with three (four) devices, \namePaper{} outperforms the best-performing prior work by 38.9\% (39.2\%) on average. 
\namePaper{}'s performance benefits come with low latency (240 $ns$ for inference) and storage (206 KiB in DRAM for both RL agents combined) overheads.
\end{abstract}

\maketitle
\thispagestyle{firstpage}
\fancyfoot[C]{\large\thepage}
\section{Introduction \label{sec:introduction}}
Modern applications (e.g., machine learning~\cite{DLRMFacebook,szegedy2017inception,ho2020denoising, ankit2019puma, low.vldb12, wang2020survey, sheng2023flexgen,pan2024instattention,jang2025inf, boroumand2021google, he2025papi, gu2025pim, park2024attacc, lincoln-hpca, lee2025aif, kim2023optimstore, LSTM, reinforce1992, gomez2022machine, qlearning_ML_1992, wang2024beacongnn, yu2024cambricon, pan2024instinfer, touvron2023llama,jiang2024mixtral,liu2024deepseek, chen2025reis}, databases~\cite{FastBitA9, redis-bitmaps, doty1980magnetic, elmasri2007fundamentals, kocberber2013meet, augusta-sigmod-2015, Oracle, idreos2012monetdb, wu2014q100, chan-sigmod-1998, oneil-ideas-2007, li-vldb-2014, li-sigmod-2013, goodwin-SIGIR-2017, seshadri-micro-2013, seshadri2017ambit, seshadri-ieeecal-2015, hajinazar2021simdram, wu-icde-1998, guz-ndp-2014, park2022flash}, graph processing~\cite{shim2022gp3d, zhang2018graphp, ahn-isca-2015, zhuo-micro-2019, huang2020heterogeneous, song2018graphr, dai-tcad-2019, beamer-SC-2012, besta2021sisa, li-dac-2016, hajinazar2021simdram}, and genome analysis~\cite{lander-nature-2001, altschul-jmb-1990, myers-jacm-1999, alser-bioinformatics-2017, loving-bioinformatics-2014, xin-bioinformatics-2015, cali-micro-2020, kim-genomics-2018, ghiasi_megis_2024, ghiasi2022genstore, papageorgiou2018genomic, soysal2025mars, alkan.naturegenetics09, ghiasi2026sage, mao2022genpip}) operate on large volumes of data, and generate irregular and rapidly evolving data access patterns that place high performance, capacity, and cost demands on storage systems.
To meet these demands, high performance computing (HPC) environments rely on hybrid storage systems (HSS) (e.g., ~\cite{xiao2016hs,wang2017larger,lee2014mining,felter2011reliability,bu2012optimization,krish2016efficient,lin2011hot,niu2018hybrid, tai2015sla, ou2014edm,cheng2015amc,zong2014faststor, matsui2017design, heuristics_hyrbid_hystor_sc_2011, lv2013probabilistic,guerra2011cost,elnably2012efficient,heuristics_usenix_2014, zhang2010automated, wu2012data, iliadis2015exaplan, lv2013hotness, matsui2017tri, feng2014hdstore, yang2017autotiering, kim2011hybridstore, liu2019hierarchical,luo2020optimal, singh2022sibyl, doudali2019kleio,ren2019archivist,cheng2019optimizing,shetti2019machine, sun2013high,              li2017utility,agarwal2015page_HMM,agarwal2017thermostat_HMM,ham2013disintegrated, salkhordeh2015operating, pavlovic2013data, meza2013case, chou2017batman, wang2019panthera, ramos2011page, doudali2021cori,sen2019machine, wu2019data, yang2014statistics, yolchuyev2021data, strunk2012hybrid, xing2025proactive,  lin2017efficient, lin2018buffer, zhang2010adaptive,   vengerov2008reinforcement,vasilakis2020hybrid2, qasem2026optimizing}),
which integrate multiple heterogeneous storage devices with diverse latency, bandwidth, endurance, and capacity characteristics (e.g., fast solid-state drives (SSDs)~\cite{inteloptane,intels4510,samsung2017znand, samsung-980pro, samsungmlc, adatasu630,intelqlc, intelp4610, nadig2023venice, cai-procieee-2017, micheloni-insidenand-2010, micheloni2013inside, cai-date-2012, cai-date-2013, cai-dsn-2015, cai-hpca-2015, cai-hpca-2017, cai-iccd-2012, cai-iccd-2013, cai-insidessd-2018, cai-sigmetrics-2014, luo2015warm, park-asplos-2021, nadig2026conduit} with small capacities and slow hard disk drives (HDDs) with large capacities~\cite{seagate, deng2011future, acharya-asplos-1998, riedel-computer-2001, keeton-sigmod-1998}.

The performance of an HSS is highly dependent on two key policies:
(1) the \emph{data-placement} policy, which determines the most suitable device in the HSS for each incoming write or allocation request, and 
(2) the \emph{data-migration} policy, which reorganizes data across the devices (i.e., prefetch hot data to the faster device and evict cold data to slower devices) to sustain high HSS performance, which can degrade over time due to (i) data misplacement, and (ii) changes in workload access patterns and HSS conditions.

\sloppypar{Designing effective placement and migration policies for modern HPC environments is \emph{non-trivial} for four key reasons.
First, HPC workloads exhibit rapidly changing access patterns (e.g., bursty checkpoint writes, multi-tenant accesses) and cause frequent changes in HSS conditions (e.g., device capacity utilization, read/write latency variations).
Second, the data-placement policy operates on the critical path of I/O handling, where \cami{microsecond-scale latency variations} directly impact application performance. Hence, placement decisions must be accurate and incur low performance overhead, even under high concurrency.
Third, poorly timed data migrations (i) interfere with \cami{incoming I/O requests from the application}, (ii) cause frequently accessed data to remain on slower devices, and (iii) exacerbate fast device contention or underutilization, which degrades throughput and tail latency.
Fourth, placement and migration policies operate in a shared, dynamic environment. Without coordination, they may make conflicting decisions that adversely impact HSS performance and device lifetimes.}

\head{Limitations of Prior Works}
Prior HSS data-management works optimize either data placement (e.g.,~\cite{matsui2017design,sun2013high,heuristics_hyrbid_hystor_sc_2011, lv2013probabilistic, guerra2011cost, elnably2012efficient, heuristics_usenix_2014, bu2012optimization, krish2016efficient, tai2015sla, zhang2010automated, wu2012data, iliadis2015exaplan, lv2013hotness, matsui2017tri, feng2014hdstore, yang2017autotiering, singh2022sibyl, doudali2019kleio,ren2019archivist,cheng2019optimizing,shetti2019machine,li2017utility,agarwal2015page_HMM,agarwal2017thermostat_HMM,ham2013disintegrated, salkhordeh2015operating, pavlovic2013data, meza2013case, chou2017batman, kim2011hybridstore, wang2019panthera, ramos2011page, liu2019hierarchical,luo2020optimal,doudali2021cori,sen2019machine, wu2019data, yang2014statistics, yolchuyev2021data, strunk2012hybrid, xing2025proactive}) or data migration (e.g.,~\cite{doudali2019kleio, lin2011hot, lin2017efficient, lin2018buffer, zhang2010adaptive, cheng2015amc, shetti2019machine, vengerov2008reinforcement,vasilakis2020hybrid2, qasem2026optimizing}) in isolation, leading to suboptimal performance in real deployments.
This is because an optimized placement policy cannot sustain long-term performance if the migration decisions remain static or oblivious to changes in workload and HSS conditions, while an optimized migration policy cannot compensate for \cami{suboptimal} placement that causes unnecessary data movement, wasted bandwidth, and increased device wear.
Our motivational study on performance- (cost-) optimized HSS configurations shows that even the state-of-the-art \cami{data-placement technique}, \sibyl~\cite{singh2022sibyl}, 
\camii{incurs 100.4\% (176.2\%) higher average I/O request latency compared to an \emph{\oracle{}}}\cami{, where Oracle represents an ideal policy with perfect knowledge of future I/O access patterns that makes optimal placement and migration decisions} (see \S\ref{sec:motivation_limitations}). \camiii{\sibyl{}'s higher latency is because of its heuristic migration policy that (1) relies on limited workload features (e.g., page access frequency), and (2) migrates pages on the critical path of I/O handling.} 

\head{Extended Data-Management Techniques}
Since no prior work jointly optimizes \emph{both} placement and migration, we construct four extended techniques (see \S\ref{sec:motivation_limitations}) that combine state-of-the-art \cami{data} placement (e.g., Sibyl~\cite{singh2022sibyl}) and migration (e.g., RNN-HSS~\cite{doudali2019kleio}) policies.
Our evaluation in \S\ref{subsec:motivation_effectiveness} shows that these techniques \cami{perform worse than} their constituent techniques because the placement and migration policies are oblivious to each other's actions, which leads to conflicting decisions, unnecessary data movement, wasted bandwidth, reduced device lifetimes, and lower end-to-end performance.
These results motivate the need for coordinated design of placement and migration policies to achieve high HSS performance.

\camii{\textbf{Our goal} is to design a holistic HSS data-management technique that jointly optimizes \emph{both} placement and migration policies in a coordinated manner to improve overall HSS performance.}
To this end, we propose \textit{\textbf{\namePaper{}}},\footnote{\namePaper{}~\cite{wiki:harmonia}: Greek goddess of harmony and balance, who brings peace to conflicts.} a multi-agent online reinforcement learning (RL) technique that co-optimizes both policies while achieving coordination between them.

\head{Key Idea} \namePaper{} employs two
lightweight autonomous RL agents, a data-placement agent and a data-migration agent, that coordinate to improve HSS performance.
\namePaper{}'s data-placement agent identifies the most suitable HSS storage device to place incoming pages. 
It builds upon the state-of-the-art \cami{data} placement technique, \sibyl~\cite{singh2022sibyl}, but incorporates (1) new \cami{migration-related} state features, and (2) a reward structure that encourages effective utilization of the faster storage devices in the HSS (see \S\ref{subsec:mechanism_rl_formulation}).
\namePaper{}'s data-migration agent continuously monitors previously-placed pages in HSS, and identifies migration candidates and their target devices. \namePaper{} assigns rewards to the migration agent based on the impact of its migrations on future HSS performance.

While both agents operate independently, they coordinate with each other through actions that affect the shared HSS environment. 
This decoupled multi-agent design enables each agent to specialize in its task while mitigating policy interference (e.g., ~\cite{yang2020multi,yu2020meta,teh2017distral,sun2020adashare}) inherent in a single-agent RL formulation.
We demonstrate in \S\ref{subsec:motivation_effectiveness} that 
a single agent that attempts to optimize both placement and migration incurs \camii{21.9\% (30.5\%) \emph{higher} latency than the best prior approach} on \cami{a} performance- (cost-) optimized HSS due to task interference: optimizing migration interferes with placement policy, as they are two different tasks (See \S\ref{subsec:needmarl}). 

\head{Evaluation} We evaluate \namePaper{} on a \emph{real Linux system} using 25 data-intensive workloads and various HSS configurations with up to four devices. 
Compared to the best-performing prior approach, Sibyl~\cite{singh2022sibyl}, \namePaper{} improves average \camii{I/O request latency} by (1) 29.3\% (44.8\%) on \cami{a} performance- (cost-) optimized HSS with two storage devices, and (2) 38.9\% (39.2\%) on HSS with three (four) devices. \namePaper{} incurs low latency (240 $ns$ for inference) and storage (206 KiB for both agents together) overheads, which makes it practical for latency-sensitive HPC applications.

This work makes the following key contributions:
\begin{itemize}[leftmargin=*]
\item \cami{We show that prior \camii{hybrid storage system (HSS)} data-management approaches \emph{do not} optimize both placement and migration. We empirically demonstrate their limitations on a real Linux system across diverse HSS configurations.}
\item We construct four extended techniques by combining state-of-the-art data placement and migration policies, and show that they \cami{perform worse than} their constituent techniques. 
\item We propose \namePaper{}, a lightweight multi-agent RL-based HSS data-management technique that co-optimizes placement and migration policies using two RL agents. 
\item Via a rigorous real-system evaluation on various HSS configurations, we demonstrate that \namePaper{} consistently outperforms state-of-the-art data-management techniques across a wide range of data-intensive workloads.
\end{itemize}
\section{Background\label{sec:background}}

\subsection{Hybrid Storage Systems}
\label{subsec:HSS-Background}
\begin{figure}[h]
\centering
\includegraphics[width=\linewidth]{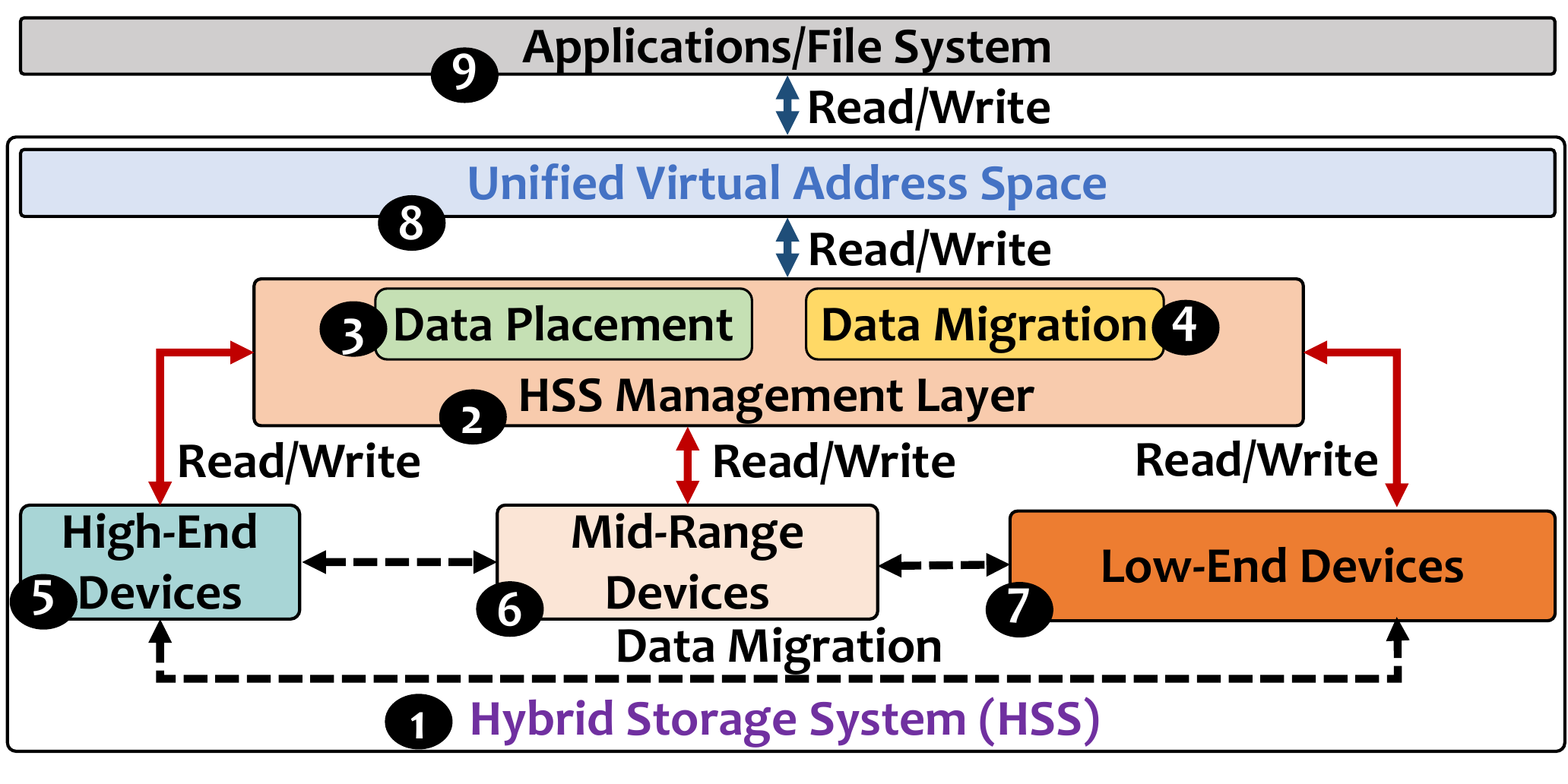}
\vspace{-2em}
\caption{Overview of a hybrid storage system.}
\label{fig:HSS-background}
\end{figure}

\fig{\ref{fig:HSS-background}} presents an overview of a hybrid storage system (HSS) (e.g., ~\cite{xiao2016hs,wang2017larger,lee2014mining,felter2011reliability,bu2012optimization,krish2016efficient,lin2011hot,niu2018hybrid, tai2015sla, ou2014edm,cheng2015amc,zong2014faststor, matsui2017design, heuristics_hyrbid_hystor_sc_2011, lv2013probabilistic,guerra2011cost,elnably2012efficient,heuristics_usenix_2014, zhang2010automated, wu2012data, iliadis2015exaplan, lv2013hotness, matsui2017tri, feng2014hdstore, yang2017autotiering, kim2011hybridstore, liu2019hierarchical,luo2020optimal, singh2022sibyl, doudali2019kleio,ren2019archivist,cheng2019optimizing,shetti2019machine, sun2013high,              li2017utility,agarwal2015page_HMM,agarwal2017thermostat_HMM,ham2013disintegrated, salkhordeh2015operating, pavlovic2013data, meza2013case, chou2017batman, wang2019panthera, ramos2011page, doudali2021cori,sen2019machine, wu2019data, yang2014statistics, yolchuyev2021data, strunk2012hybrid, xing2025proactive,  lin2017efficient, lin2018buffer, zhang2010adaptive,   vengerov2008reinforcement,vasilakis2020hybrid2, qasem2026optimizing})~\circled{1}. 
An HSS has two key components: (1) multiple storage devices with diverse characteristics (e.g., I/O latency, I/O bandwidth, and device capacity) and storage protocols (e.g., NVM Express (NVMe)~\cite{nvme} or SATA~\cite{sata}), and (2) an HSS management layer~\cite{singh2022sibyl}~\circled{2} that orchestrates data placement (e.g.,~\cite{matsui2017design,sun2013high,heuristics_hyrbid_hystor_sc_2011, lv2013probabilistic, guerra2011cost, elnably2012efficient, heuristics_usenix_2014, bu2012optimization, krish2016efficient, tai2015sla, zhang2010automated, wu2012data, iliadis2015exaplan, lv2013hotness, matsui2017tri, feng2014hdstore, yang2017autotiering, singh2022sibyl, doudali2019kleio,ren2019archivist,cheng2019optimizing,shetti2019machine,li2017utility,agarwal2015page_HMM,agarwal2017thermostat_HMM,ham2013disintegrated, salkhordeh2015operating, pavlovic2013data, meza2013case, chou2017batman, kim2011hybridstore, wang2019panthera, ramos2011page, liu2019hierarchical,luo2020optimal,doudali2021cori,sen2019machine, wu2019data, yang2014statistics, yolchuyev2021data, strunk2012hybrid, xing2025proactive})~\circled{3} and data migration (e.g.,~\cite{doudali2019kleio, lin2011hot, lin2017efficient, lin2018buffer, zhang2010adaptive, cheng2015amc, shetti2019machine, vengerov2008reinforcement,vasilakis2020hybrid2, qasem2026optimizing})~\circled{4} across the HSS devices.
A typical HSS combines fast, low-capacity devices (i.e., high-end device~\cite{inteloptane,intels4510,samsung2017znand, samsung-980pro, samsungmlc, adatasu630,intelqlc, intelp4610, nadig2023venice, cai-procieee-2017, micheloni-insidenand-2010, micheloni2013inside, cai-date-2012, cai-date-2013, cai-dsn-2015, cai-hpca-2015, cai-hpca-2017, cai-iccd-2012, cai-iccd-2013, cai-insidessd-2018, cai-sigmetrics-2014, luo2015warm, park-asplos-2021})~\circled{5}) and slower, high-capacity devices~\cite{seagate, deng2011future, acharya-asplos-1998, riedel-computer-2001, keeton-sigmod-1998} (i.e., mid-range~\circled{6} and low-end devices~\circled{7}) to balance performance, capacity, and cost efficiency for deployment in HPC storage nodes.

\noindent \head{HSS Management Layer} 
The HSS management layer, typically implemented in the host OS kernel space, is conceptually similar to the Linux \emph{md}~\cite{mdlinux} subsystem, which provides software RAID (RAID 0)~\cite{chen1994raid, patterson1988case, schulze1989reliable} across multiple devices. 
Unlike RAID, which places data across storage devices using fixed striping policies and does not perform runtime migration, the HSS management layer dynamically performs both data placement and migration.

The HSS management layer performs three key functions.
First, it 
(i) exposes a unified flat \camii{virtual address space~\circled{8}} to applications/\camii{file system}~\circled{9}, allowing data access without exposing the underlying device heterogeneity, 
(ii) maps virtual addresses to device-specific logical block addresses (LBAs),
and (iii) translates read/write operations from the applications to NVMe~\cite{nvme} or SATA~\cite{sata} I/O requests.
Second, it performs two performance-critical tasks: (1) placing incoming I/O request data (i.e., data placement~\circled{3}) in the best-fit storage device, and (2) reorganizing previously-placed data (i.e., data migration~\circled{4}) across devices to sustain high performance over time.
Third, to perform data placement and migration, it maintains metadata in the host DRAM, which includes virtual address to device-specific LBA mapping, I/O request \camii{characteristics} (e.g., request type, request size), and storage device characteristics (e.g., device capacities and current utilization).

\subsection{Reinforcement Learning}
\label{subsec:reinforcement_learning-Background}

\begin{figure}[h]
\centering    
\includegraphics[width=0.7\linewidth]{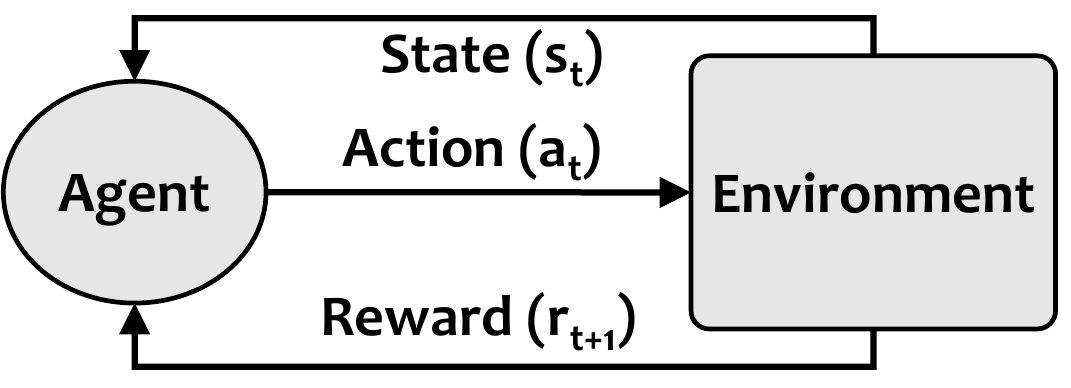}
    \vspace{-1em}
    \caption{Overview of reinforcement learning.}
    \label{fig:rl-background}
\end{figure}

Reinforcement learning (RL) \camii{(e.g., ~\cite{sutton1998reinforcement, singh2022sibyl, ipek2008self, multi_scheduler_HPCA_2012, bera2021pythia, kang2017reinforcement, ha2024rl, lu2022rlrp, li2023rlalloc, bera2026athena, bera2026mitigating, andrew2018reinforcement, bera2026machine})} is a machine learning (ML) approach in which an agent \camiv{aims to learn} the optimal policy for a specific objective by interacting with its environment. 
An RL formulation (see \fig{\ref{fig:rl-background}})  consists of four key components.

\noindent
\textbf{1. State}. The state represents the observable information about the environment at a given time. Assuming $S$ is the set of all possible states, $s_t$ $\in$ $S$ is the state at \emph{time step} $t$. 

\noindent
\textbf{2. Action}. At each time step, the agent observes the state $s_t$ and performs an action $a_t$ from the set of all actions $A$, which causes the environment to transition to a new state $s_{t+1}$.

\noindent
\textbf{3. Reward}. In response to $a_t$ that changes the environment state from $s_t$ to $s_{t+1}$, the agent receives a numerical \emph{reward} $r_{t+1}$ that quantifies the utility of the chosen action.

\noindent
\textbf{4. Policy}. The policy $\pi$ defines how the agent selects actions in each state. 
The RL agent aims to learn an optimal policy \textbf{$\pi^{*}$} that maximizes the cumulative reward over time. The optimal policy is determined by computing the optimal action-value function $Q^{*}$, also called the Q-value. The Q-value of a state-action pair, denoted as $Q(s, a)$, represents the expected cumulative reward obtained by taking action $a$ in state $s$.

\head{Multi-Agent RL (MARL)} 
MARL (e.g., \cite{canese2021multi, jain2017coordinated, qiu2022reinforcement, qiu2022simppo, qiu2020firm, mao2023multi, mlsys2023qiu, lu2022rlrp, jain2017cooperative, shen2022multiagent}) involves multiple RL agents interacting with a shared environment and with one another. 
Prior works~\cite{yang2020multi,yu2020meta,teh2017distral,sun2020adashare, ammar2014automated, carroll2005task} show that a single RL agent struggles to effectively optimize multiple tasks with low task similarity (i.e., different inputs, objectives, and actions) because learning from one task can interfere with learning for another. This interference leads to unstable training and suboptimal policies.
MARL can effectively handle multiple dissimilar tasks by enabling each agent to specialize in its task with its own objective and timescale.
\section{Motivation}\label{sec:motivation_limitations}
\subsection{Why RL for HSS Data Management? \label{subsec:whyRL}}
\noindent RL is a good fit for HSS data management for four key reasons. 

First, RL learns online from system-level feedback without (i) prior training on labeled data, or (ii) frequent retraining. 
This enables continuous adaptation to changes in workload access patterns and system conditions, which are common in HSS environments. 

Second, RL naturally handles complex decision-making tasks in which each action modifies the system state and influences long-term system behavior. 
By optimizing cumulative returns (e.g.,~\cite{sutton1998reinforcement,sutton1999policy, ke2018modeling}), RL learns policies that account for long-term effects of their decisions. 
This property makes RL well-suited for HSS data management where placement and migration decisions affect device utilization (e.g., bandwidth) and overall system performance.

\begin{figure}[h]
\centering    
\includegraphics[width=\linewidth]   {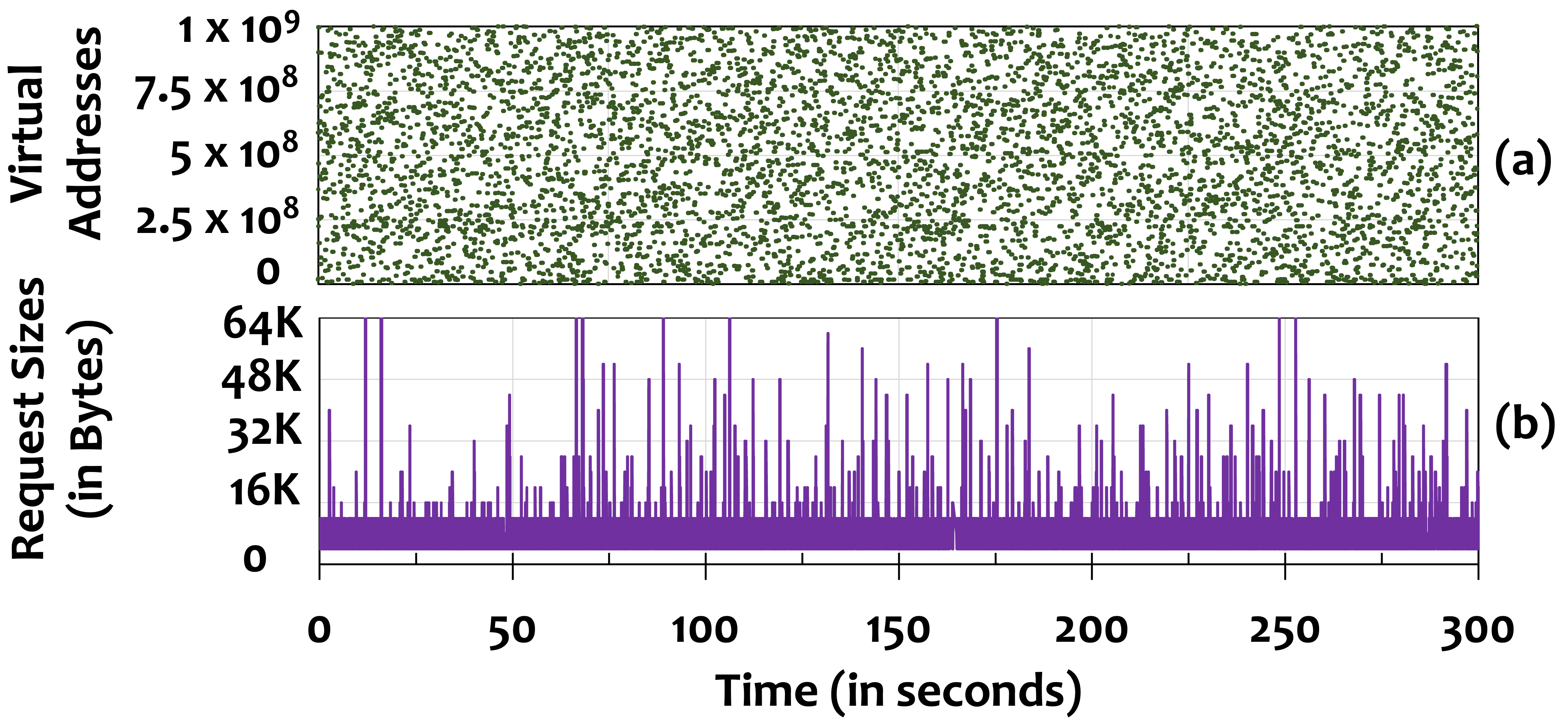}
\vspace{-2em}
\caption{Virtual addresses accessed (a)  and request sizes (b) captured during a 300-second execution of ssd-02 workload. Each virtual address accessed is represented as a green dot.}    
\label{fig:workload_characteristics}
\end{figure}

Third, RL is effective even for irregular access patterns typical of HSS environments because it implicitly learns value-based policies that estimate the performance impact of decisions rather than directly predicting future I/O \camii{requests}.
\camii{To illustrate irregular access patterns in modern data-intensive workloads, we show the virtual addresses accessed and I/O request sizes during a 300-second execution of one of our chosen workloads, ssd-02 (see \S\ref{sec:evaluation_methodology} for evaluation methodology), in \figs{\ref{fig:workload_characteristics}}(a) and \ref{fig:workload_characteristics}(b) respectively.
We make two key observations: 
(1) the application accesses a wide range of virtual addresses (1$\times$$10^6$ - 1$\times$$10^9$) without recurring access patterns. 
(2) the I/O request sizes vary from random accesses (i.e., 4KiB) to large sequential accesses (i.e., \camiv{128 KiB}).
We observe similar dynamic variations in many of our chosen workloads.    
Prior heuristic and supervised learning based approaches (see \camiv{\S\ref{subsec:motivation_methodology}}) make fixed decisions at coarse intervals, and cannot adapt efficiently to the dynamic variations in workload access patterns and system conditions (see \S\ref{subsec:motivation_effectiveness}).
Sibyl~\cite{singh2022sibyl} shows that RL outperforms heuristic-based (e.g.,~\cite{matsui2017design, matsui2017tri}) and supervised-learning-based (e.g., ~\cite{doudali2019kleio, shetti2019machine}) approaches for HSS data placement.
} 

Fourth, RL readily generalizes across diverse system configurations with minimal \camii{additional} designer effort~\cite{mao2019learning, mirhoseini2021chip, mao2016resource, singh2022sibyl}. This flexibility makes RL suitable for HSS configurations that integrate multiple storage devices with heterogeneous performance and capacity characteristics (as partially demonstrated in Sibyl~\cite{singh2022sibyl}). 

\subsection{Need for Multi-Agent RL for HSS \label{subsec:needmarl}}
Although RL is well-suited for HSS data management, a single RL agent is \emph{not} ideal for jointly optimizing placement and migration in HSS, as these two tasks differ fundamentally in three key ways.
First, they operate on different data. Data placement decides where \camii{\emph{new I/O request data}} should be written. In contrast, data migration reorganizes \camii{\emph{previously-placed data}} across devices.
Second, their objectives differ. Placement aims to minimize per-request I/O latency \camii{and manage fast-device utilization on the critical path}, but migration focuses on long-term HSS performance.
Third, these two tasks operate on different timescales. Placement is on the critical path of I/O handling. Migration operates asynchronously to the application’s I/O requests.

These differences cause interference in a single agent's learning when \camii{the agent} attempts to jointly optimize both tasks (see \S\ref{subsec:reinforcement_learning-Background}). As we demonstrate in \S\ref{subsec:motivation_effectiveness}, a single-agent RL formulation (SAPM) performs poorly when handling both data placement and migration in HSS, motivating a multi-agent RL formulation.
MARL allows each agent to specialize in its task \camii{(i.e., learn a policy tailored to its own objective, decision timescale, and state features)} while coordinating to achieve system-level objectives. 
MARL suits HSS data management, where latency-critical placement decisions and longer-term migration decisions must be jointly optimized without interfering with each other.

\subsection{Prior HSS Data Management Techniques\label{subsec:motivation_methodology}}

Prior HSS data-management techniques typically optimize either data placement or data migration in isolation. We describe representative state-of-the-art data placement and migration techniques. 

\head{Data-Placement Techniques}
We evaluate two state-of-the-art HSS data-placement techniques: \cde{} (\underline{\textbf{C}}old \underline{\textbf{D}}ata \underline{\textbf{E}}viction)~\cite{matsui2017design} and \sibyl{}~\cite{singh2022sibyl}.
\camii{We select these techniques because they represent two widely-used classes of prior HSS data-placement policies: heuristic-based placement (\cde{}) and ML-based placement (\sibyl{}).}
\cde{} is a heuristic placement technique that places hot or random (cold or sequential) data in the fast (slow) device. \cde{} determines hotness and randomness of data, using past access frequency and request size, respectively.
\sibyl{}~\cite{singh2022sibyl} uses an RL agent that considers multiple features from I/O requests (e.g., request type and size) and the HSS state (e.g., device capacity, I/O latency) to make placement decisions.   
When the fast device is full, both techniques evict least-recently-used (LRU) data from the fast device before placing new I/O requests.

\head{Data-Migration Techniques} 
We evaluate two state-of-the-art HSS data-migration techniques, \ksvm{}~\cite{shetti2019machine} and \rnn{}~\cite{doudali2019kleio}.
\camii{These techniques represent strong learning-based migration policies that proactively reorganize data in HSS based on predicted page access frequency.}
Both use supervised learning (e.g., ~\cite{giles1994dynamic, pearlmutter1989learning, noble2006support, hearst1998support, sutskever2014sequence}) to identify the target device for migration based on page access frequency. 
\ksvm{} uses a k-means-assisted support vector machine (SVM) classifier~\cite{hearst1998support, noble2006support} trained on past access frequencies to identify migration candidate pages and their target devices; 
\rnn{} uses \camii{a} recurrent neural \camii{network} (RNN)~\cite{giles1994dynamic, pearlmutter1989learning, sutskever2014sequence} to predict future page access frequencies. Using these predictions, \rnn{} identifies migration candidate pages and their target devices.
Both perform migrations at fixed intervals (e.g., after every 1000 incoming requests) on the critical path of I/O request handling.
Both techniques use a fixed heuristic placement policy (e.g., \ksvm{}~\cite{shetti2019machine} places all data in the fast device until it is full).

\head{Extended Data-Management Techniques}
Since \emph{no} prior technique optimizes \camii{\emph{both}} data placement and migration, we construct four extended data-management techniques to evaluate whether combining the state-of-the-art data placement and migration policies improves performance. 
\sibyl{}+\ksvm{} and \sibyl{}+\rnn{} combine \sibyl{}, the state-of-the-art data placement technique, with prior ML-based migration techniques.
\cde{}+\rlmigr{}, combines \cde{}, a heuristic data placement technique, with an RL-based migration policy that uses the same design (i.e., state features and reward structure) as \namePaper{}'s migration agent \camiv{(see \S\ref{subsec:mechanism_rl_formulation})}.
\sapm{} (Single Agent for Placement and Migration) explores whether a single RL agent can jointly optimize placement and migration. \sapm{} performs placement for every incoming I/O request, and migrates data only during system idle times to avoid interference with placement. We use the same state features and reward structure as \namePaper{} \camiv{(see \S\ref{subsec:mechanism_rl_formulation})} in \sapm{}, and we carefully tune its hyperparameters to achieve the best performance possible with a single-agent formulation.
\camii{
Note that \cde{}+\rlmigr{} and \sapm{} are new techniques we develop in this work, in order to develop an even better technique (\namePaper{}).}

\subsection{Effectiveness of Prior Techniques \label{subsec:motivation_effectiveness}}

\begin{figure}[h]
\centering
\includegraphics[width=\linewidth]{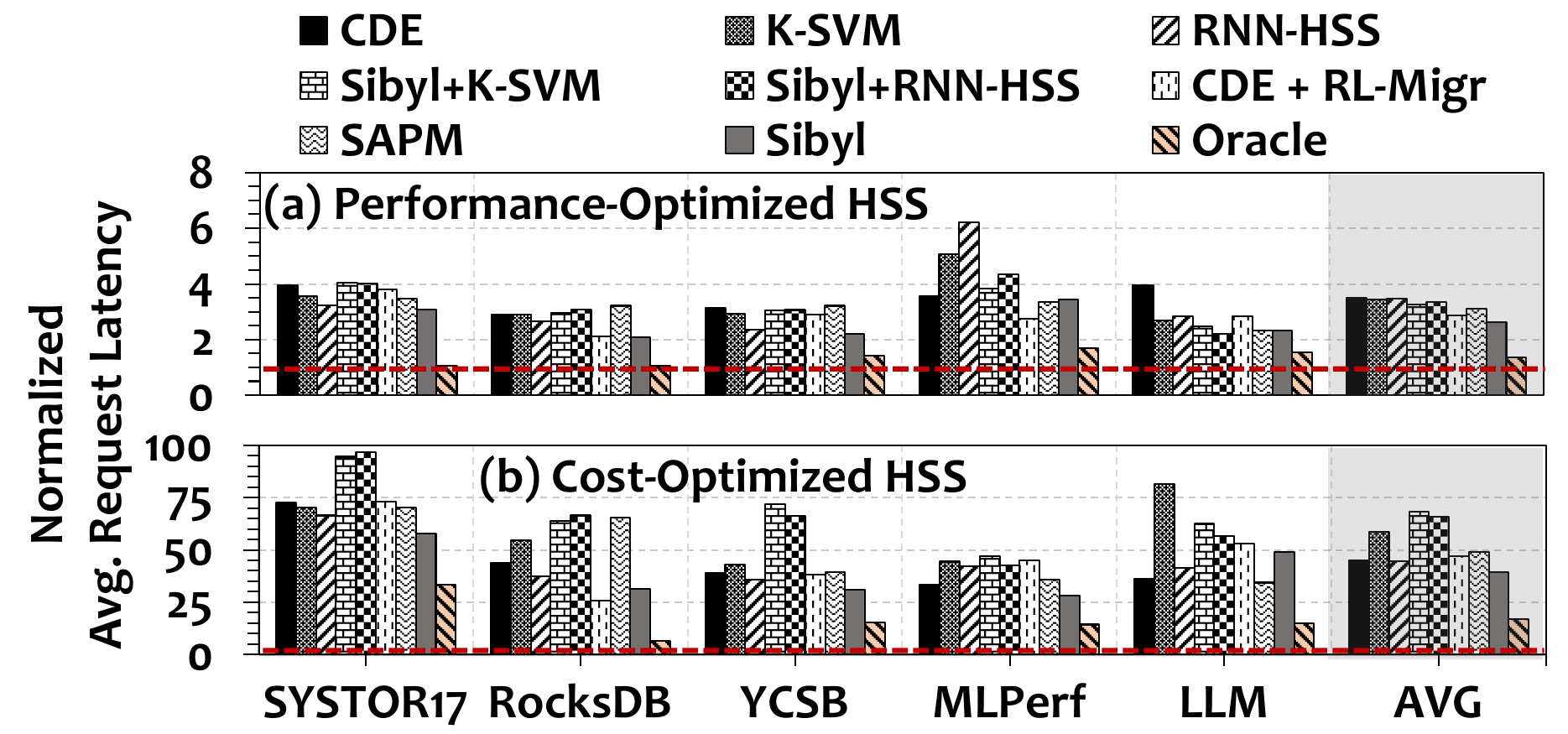}
\vspace{-2em}
\caption{\camii{Average I/O latency of \cde{}, \ksvm{}, \rnn{}, \sibyl{}, \sibyl{}+\ksvm{}, \sibyl{}+\rnn{}, \cde{}+\rlmigr{}, \sapm{} and \oracle{} on a performance- (cost-) optimized HSS, normalized to the latency of Fast-Only. Lower is better.}}
\label{fig:motivation}
\end{figure}

\head{Methodology} 
We evaluate eight prior data-management techniques on a real system \camii{(see Table \ref{tab:devices})} with performance- and cost-optimized HSS configurations across 20 data-intensive workloads from five benchmark suites (see Table \ref{tab:workloads}). \S\ref{sec:evaluation_methodology} describes our evaluation methodology.
We compare the \camii{average I/O request latency} of these \camii{eight} techniques against two \emph{ideal} baselines: \camii{\emph{Fast-Only}} and \camii{\emph{\oracle{}}}.
Fast-Only is an ideal policy \camii{where} the fast storage device is large enough to accommodate the entire workload.
\oracle{}~\cite{meswani2015heterogeneous} represents an upper bound that \camii{minimizes average I/O request latency by making data placement and migration decisions} with complete knowledge of the \camii{\emph{future}} I/O access patterns of the entire workload, while respecting device-capacity constraints. 
\oracle{} performs migration during idle periods, and does \camii{\emph{not}} incur migration-induced latency overhead.
All values in \fig{\ref{fig:motivation}} are normalized to Fast-Only.

\head{Performance Results}
\fig{\ref{fig:motivation}} shows the \camii{average I/O request latency} of four prior and four extended data-management techniques on (a) a performance-optimized and (b) a cost-optimized HSS. We make five key observations. 

First, \camii{\emph{all}} prior techniques fall short of \oracle{} for \emph{every} HSS configuration and workload. 
The best-performing \camii{prior} approach, \sibyl{}, \camii{incurs 100.4\% (176.2\%) higher average I/O latency} than \oracle{} in a performance- (cost-) optimized HSS.
While \sibyl{}'s RL-based placement adapts to workload and system changes, its heuristic migration policy (i) cannot adapt to changes in workload and system conditions, and (ii) \camii{evicts pages} on the critical path of I/O handling.

Second, \sibyl{} outperforms \cde{}, \ksvm{}, and \rnn{} by 24.7\% (9.7\%), 22.4\% (32.9\%), and 19.0\% (11.4\%), respectively, on a performance- (cost-) optimized HSS due to its adaptive RL-based placement policy. In contrast, the other baselines rely on heuristic or supervised-learning approaches that use limited workload and HSS features and cannot effectively adapt to changes in workload and HSS conditions. 

Third, \sibyl{}+\ksvm{} and \sibyl{}+\rnn{} perform worse than \sibyl{} alone, as naive combinations of placement and migration policies lead to (i) unnecessary data movement due to \camii{\emph{conflicting}} placement and migration decisions, and \camiv{(ii)} \camii{\emph{slower convergence}} of \sibyl{}'s policy.

Fourth, \cde{}+\rlmigr{} \cami{performs worse than} \sibyl{} by 7.1\% (12.4\%) on average in a performance- (cost-) optimized HSS. 
It achieves \camii{latency} comparable to \sibyl{} in read-intensive workloads (e.g., RocksDB, MLPerf) because \cde{}+\rlmigr{}'s RL-based migration agent proactively prefetches frequently read data \camii{into the faster device}. \cde{}+\rlmigr{} performs worse in write-intensive workloads (e.g., \camii{SYSTOR17}) because its heuristic placement policy \camiii{places all data in the fast device and} triggers frequent migrations.
This result indicates that an adaptive migration policy requires an equally adaptive placement policy to \camii{provide} consistent performance gains.

Fifth, \sapm{} \cami{performs worse than} \sibyl{} by \camii{21.9\% (30.5\%)} on average in the performance- (cost-) optimized HSS due to two key reasons: (1) \sapm{} cannot effectively optimize placement and migration tasks due to their differing objectives and timescales (see \S\ref{subsec:needmarl}), and (2) \sapm{} focuses on migration only when there are no placement requests, which leads to untimely migrations.

These results demonstrate that existing techniques \camii{(and new techniques that we develop to address the problem with a single RL agent)} optimize either placement or migration in isolation, and \camiv{naive} combinations of state-of-the-art data placement and migration techniques remain insufficient.

\textbf{Our goal} is to design a holistic HSS data-management technique that jointly optimizes \camii{\emph{both}} placement and migration policies \camii{in a coordinated manner} to improve overall HSS performance.
\section{\namePaper{}\label{sec:mechanism}}
We design \camii{\namePaper{},} an online multi-agent RL \camii{(MARL)} technique for HSS data management with two autonomous RL agents: a data-placement agent and a data-migration agent. We implement \namePaper{} in the HSS management layer in the host OS.

\subsection{Key Design Challenges \label{subsec:mechanism_challenges}}
Designing a MARL-based HSS data-management technique introduces three key challenges.
First, data placement and data migration are different (see \S\ref{subsec:needmarl}) yet interdependent tasks. 
Aggressive placement decisions can trigger unnecessary migration, while delayed migrations can limit the effectiveness of placement decisions. Without careful coordination, independently learning agents may compete rather than cooperate, which \camii{can lead} to unnecessary data movement and unstable behavior.
Second, workload access patterns and device conditions evolve continuously, which makes the environment non-stationary. Multiple agents exacerbate this non-stationarity because each agent's decisions modify the environment observed by the other \camii{agent(s)}, making convergence of RL agents challenging. 
Third, multiple RL agents can increase computational complexity and storage overhead (see \S\ref{subsec:overhead_analysis}). Since placement operates on the critical path of I/O handling, the design must minimize inference latency and memory footprint. 

While MARL has been studied in other domains (e.g.,~\cite{jain2017coordinated, qiu2022reinforcement, qiu2022simppo, qiu2020firm, mao2023multi, mlsys2023qiu, lu2022rlrp, jain2017cooperative, shen2022multiagent}), these techniques use (1) identical RL agents with a \camii{\emph{single}} objective, \camii{and/or} (2) joint-action space and shared reward functions for \camii{\emph{all}} agents. 
These approaches do \camii{\emph{not}} hold for HSS data management, where placement and migration have \camii{\emph{different}} objectives, constraints, and timescales. 

\subsection{\namePaper{}: RL Formulation \label{subsec:mechanism_rl_formulation}}
\begin{figure}[h]
\centering
\includegraphics[width=0.9\linewidth]{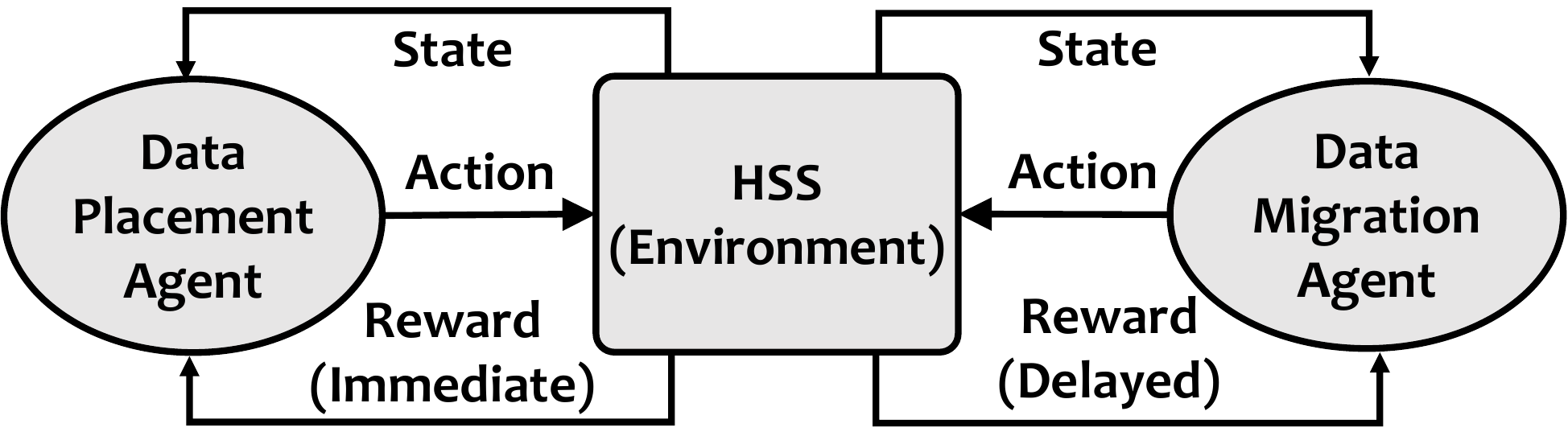}
\vspace{-1em}
\caption{RL Agents in \namePaper{}.}
\label{fig:rl-agents}
\end{figure}

\fig{\ref{fig:rl-agents}} shows \namePaper{}'s MARL formulation. \namePaper{} consists of two autonomous agents that interact with the environment (HSS) by observing state features, performing actions, and receiving rewards.
We design the state, action, and reward for both agents to enable coordination and maximize HSS performance.

\head{State}
\namePaper{}'s RL agents operate on a shared state representation consisting of seven features that capture current HSS conditions and workload characteristics. 
\camii{These features capture key factors that influence placement and migration decisions, including I/O request characteristics, current device state, and a page's access and migration history.
To minimize runtime overhead (i.e., inference latency and metadata storage) while preserving useful state information, we (1) select a small set of highly relevant features using feature selection~\cite{kira1992featureselection}, and  
(2) quantize each feature into a small number of bins (Table~\ref{tab:state}) based on workload characterization.}

At time step $t$, the state is represented \camii{using} a seven-dimensional observation vector in Equation 1. 
\begin{multline}
O_t = (req\_size_t, req\_type_t, acc\_intr_t, acc\_freq_t, \\fast\_cap_t, curr\_dev_t, migr\_intr_t)
\end{multline}
The observation vector size directly affects \namePaper{}'s storage and computational overheads (see \S\ref{subsec:overhead_analysis}). 

Table~\ref{tab:state} describes the seven state features in the observation vector.
\begin{table}[h]
\caption{State features used by \namePaper{}'s RL Agents}
\label{tab:state}
\vspace{-1em}
\centering
\renewcommand{\arraystretch}{1}
\setlength{\tabcolsep}{2pt}
\resizebox{\linewidth}{!}{%
\begin{tabular}{|c|c|c|c|}
\hline
\textbf{Feature} & \textbf{Description} & \textbf{\# of bins} & 
\textbf{Encoding (bits)}\\ 
\hline
\hline
$req\_type$ &  Request type (read/write) & 2 & 1  \\
\hline
$req\_size$ & Request size (in pages) & 8 & 3 \\
\hline
$acc\_intr$ & {\begin{tabular}[c]{@{}c@{}} Access interval \\ of the requested page \end{tabular}} & 64 & 8 \\
\hline
$acc\_freq$ & {\begin{tabular}[c]{@{}c@{}} Access frequency \\ of the requested page \end{tabular}}  & 64 & 8   \\
\hline
$fast\_cap$   %
&  {\begin{tabular}[c]{@{}c@{}} Free space \\ in the fast storage device \end{tabular}}  & 8 & 3    \\
\hline
$curr\_dev$ & {\begin{tabular}[c]{@{}c@{}} Storage device where the \\ requested page currently resides \end{tabular}} & 2 & 1\\
\hline
$migr\_intr$ & {\begin{tabular}[c]{@{}c@{}} Migration interval of a page \end{tabular}} & 64 & 8\\
\hline
\end{tabular}
 }
\end{table}
 
Request type ($req\_type$) indicates a read or a write operation, which requires one bit. 
\camii{Request size ($req\_size$) is the size of the current I/O request represented in terms of the number of 4 KiB pages. 
We quantize request size into eight classes (e.g., 1 page ($\leq$ 4 KiB), 2 pages (4-8 KiB), 4 pages (8-16 KiB), 8 pages (16-32 KiB), 16 pages (32-64 KiB), \camiv{32 pages (64-128 KiB), 64 pages (128-256 KiB),} and $\geq$ 128 pages ($\geq$ 512 KiB)), which requires three bits.
}
Access interval ($acc\_intr$) captures the number of page accesses between two accesses to the same page, and reflects temporal locality. We encode the access interval using eight bits.
Access frequency ($acc\_freq$) records the total number of times a page is accessed. Access frequency consumes eight bits.  
Fast-device capacity ($fast\_cap$) is the remaining free space in the fast storage device, and enables capacity-aware data placement and migration decisions. We quantize the fast-device capacity into eight bins, which requires three bits.
Current device \camiv{($curr\_dev$)} indicates the storage device where the page currently resides, and informs the agents about migration opportunities. The current device depends on the number of HSS storage devices, and consumes one bit for a dual-HSS. 
Migration interval ($migr\_intr$) is the number of migrations performed between two migrations of the same page, and reflects migration stability. Similar to the access interval, the migration interval is encoded using eight bits. 

Overall, we encode the state representation using 32 bits, enabling fast feature lookups and low-latency inference.
We allocate higher precision to selected state features (e.g., acc\_intr, acc\_freq, migr\_intr) to ensure accurate representation and extensibility to new workloads and HSS configurations.
\namePaper{} maintains a metadata table in host DRAM to track the state features. To balance performance and capacity, \namePaper{} maintains only a frequently accessed subset \camii{(0.1\% of the host DRAM capacity)} of this table in the host DRAM, while the remaining metadata is flushed to the fast storage device. Both agents perform a metadata lookup to observe the current state to make placement and migration decisions.

\head{Action} 
\namePaper{}’s agents operate independently, and their action spaces depend on the number of HSS devices. 
\camii{The placement agent selects a target storage device for each incoming write request sent from the application to the HSS. Placement decisions apply only to write requests because the target storage device for a read request is determined by the read data's current location in the HSS.
The migration agent selects a target storage device to migrate a previously-placed page.}
In a two-device HSS, agents choose between fast and slow devices. We encode the action space using 4 bits to reduce complexity and enable scalability to more devices. 

\head{Reward} 
The reward structure is key to finding an optimal policy in RL. We design the reward structures of \namePaper{}'s agents based on two factors. 
First, since our goal is to improve overall HSS performance, we use I/O latency as the primary feedback signal for both agents. 
I/O latency captures the internal state of HSS storage devices (e.g., queueing delays, buffer dependencies, \camii{shared resource (e.g., NAND flash channels, PCIe lanes) contention~\cite{nadig2023venice, micheloni2013inside, micheloni2018hybrid, kim2022networked, kim2021decoupled, jaliminche2023enabling, gao2017exploiting}}, garbage collection~\cite{yang2014garbage, cai-procieee-2017, tavakkol-fast-2018, agrawal2008design, shahidi2016exploring, lee2013preemptible,jung2012taking, choi2018parallelizing,wu2016gcar,cai-insidessd-2018,cai-hpca-2017}, \camii{read/write asymmetry~\cite{wu2019hotr, cui2022improving, song2023adaptive, bae20182b, lee2023lru, papon2024enhancing, li2020maximizing, an2022avoiding, wu2012adaptive, liu2019soml, shim2019exploiting, im2010flash, an2022your, lee2008case}}, \camii{non-blocking operations~\cite{campello2015non, kim2018autossd, kim2019design, yu2014optimizing, li2022fantastic, useche2011truly, du2024pipessd}}, \camii{error handling~\cite{cai-date-2012, cai-dsn-2015, cai-hpca-2017, cai-iccd-2012, luo-dsn-2014, luo-sigmetrics-2018, luo2015warm, luo-jsac-2016, cai-date-2013, cai-hpca-2015, cai-iccd-2013, cai-insidessd-2018, cai-inteltechj-2013, cai-procieee-2017, cai-sigmetrics-2014, cai.bookchapter18.arxiv}}), making it a robust proxy for system performance.
Second, we design the reward structures to enable coordination between the two agents. 

The placement agent aims to improve the \camii{latency} of the current I/O request. 
For each placement action at time step $t$, the data-placement agent receives an immediate reward $R_{placement}$ at time step $t+1$ that is inversely proportional to the I/O request latency:
\begin{equation} \label{eq_Janus_placement_reward}
R_{placement} = \frac{1}{L_t} 
\end{equation}
where $L_t$ is I/O request latency at time step $t$.
Although placing data on the fast device yields lower latency and a higher reward, \namePaper{} does not greedily choose this action. Instead, it learns to balance device utilization because: 
(1) Fast device capacity is limited, and its overuse can degrade performance due to \camii{queueing delays, bandwidth contention, and garbage collection}.
(2) Selectively placing less-critical data on slower devices preserves fast-device capacity for latency-sensitive requests and reduces contention.

The migration agent aims to maximize long-term HSS performance because:
(1) the application may not immediately access the prefetched data, and 
(2) the performance gains from freed fast-device capacity \camiv{are} typically \camii{observed} only during subsequent placements. 
Hence, after migrating $x$ pages to \camii{the pages' identified} target devices, the migration agent receives a \emph{delayed} reward that is inversely proportional to the average data-placement latencies (i.e., \(\frac{\sum L_i}{n}\)) \camii{of} the next $n$ I/O requests. The delayed reward serves as feedback for previous migration decisions. 
\begin{equation}
\label{eq_Janus_migration_reward}
R_{migration} = \begin{cases}
        \scalemath{1}{\frac{n}{\sum_{i=t}^{t+n}L_i}}-\textit{$P_{migr}$}
        & \text{\textit{after migrating $x$ pages}} 
        \\
        0 & \textit{otherwise}
    \end{cases}
\end{equation}
where $t$ is the current time step and $L_i$ is the latency of I/O request $i$.
To avoid repeated migrations of the same page (\emph{ping-pong migrations}), we add a small penalty $P_{migr}$ inversely proportional to the average migration and access intervals (See Table \ref{tab:state}) of $x$ pages.
Based on empirical analysis (see \S\ref{subsec:perf_results}), we set $n$ to 50 and $x$ to 10, i.e., the migration queue size. 
We use a 16-bit representation for the reward structure of both agents to reduce storage overhead.

\subsection{Coordination Between \namePaper{}'s Agents \label{subsec:mechanism_coordination}} 
We enable coordination between the two agents through (1) their rewards, and (2) the impact of their actions on the HSS. 
The placement agent's reward (see Equation~\ref{eq_Janus_placement_reward}) encourages effective fast device utilization, which depends on the migration agent proactively migrating cold (hot) data to the slow (fast) device. 
The migration agent receives a delayed reward (see Equation 3) based on placement latencies of $n$ I/O requests after migrating candidate pages. This delayed feedback encourages migration decisions that improve the effectiveness of future data placement. 

\head{Mitigating Conflicts} 
To prevent conflicting placement and migration decisions, we use two approaches. 
First, the migration agent monitors pages with high access and migration intervals (see Table~\ref{tab:state}), and ignores recently placed or migrated pages \camii{(e.g., if the page was accessed in the previous request)}.
Second, we incorporate a small penalty in the migration agent's reward (Equation~\ref{eq_Janus_migration_reward}) to discourage repeated migrations of the same page (i.e., ping-pong migrations).

\begin{figure}[h]
\centering
\includegraphics[width=\linewidth] {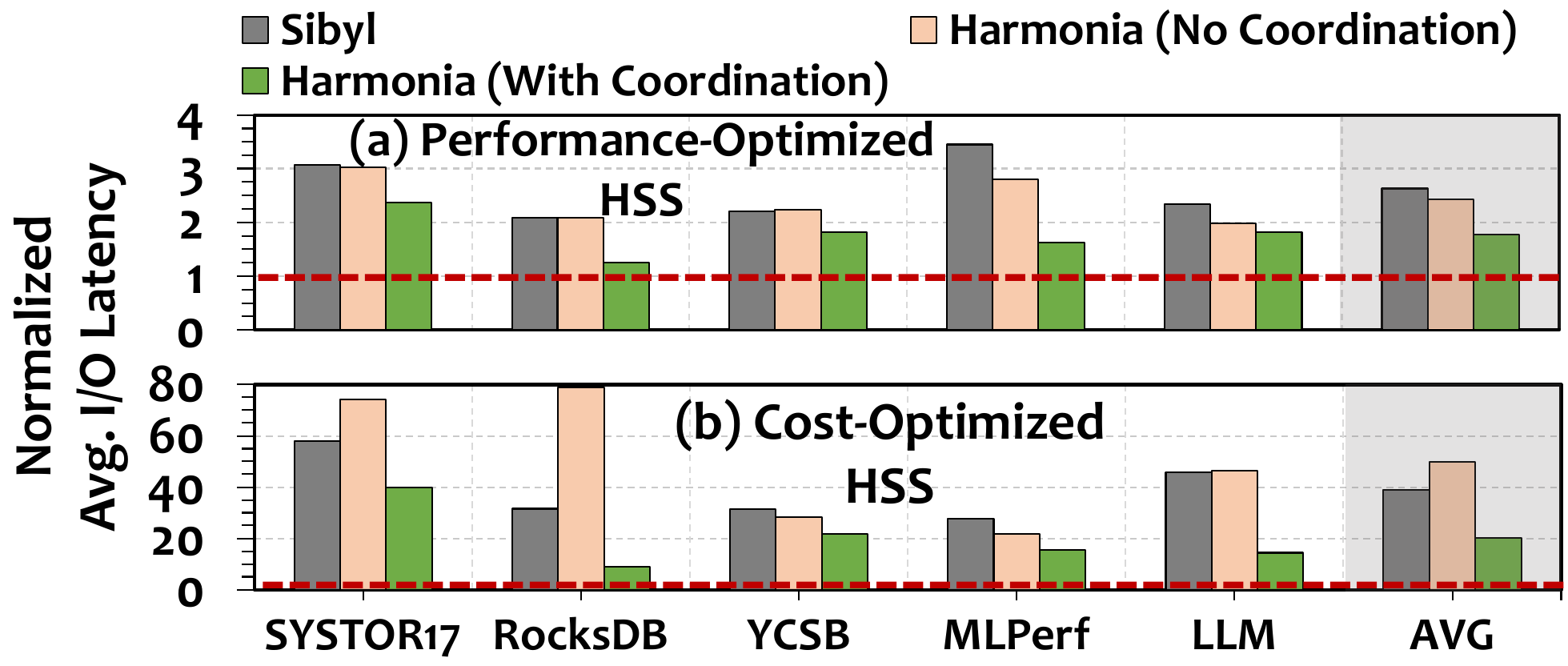}
\vspace{-2em}
\caption{\camii{Average request latency of \sibyl{}, \namePaper{} (No Coordination) and \namePaper{} (With Coordination) on a performance- (cost-) optimized HSS\camii{,} normalized to the latency of Fast-Only.}}
\label{fig:coordination_perf}
\end{figure}

\head{Impact of Coordination}
To evaluate the impact of coordination between \namePaper{}'s agents, we compare the \camii{average request latency} of \sibyl{}, \namePaper{} without coordinated agents \camii{(\namePaper{} (No Coordination))},\footnote{For \camii{\namePaper{} (No Coordination)} we choose one of several possible rewards that cause the placement and migration agents to operate \camii{\emph{independently}}. \camii{The placement agent optimizes only immediate placement latency. The migration agent minimizes the latency cost of page migrations. Each agent makes decisions \camii{\emph{without}} considering how its actions affect the other agent's decisions.} This naive MARL formulation \camii{eliminates} reward-based coordination between agents.} and \namePaper{} with \camiv{Coordination}\footnote{\namePaper{} with \camiv{Coordination} is the complete version of our proposed technique\camii{, as fully described in \S\ref{subsec:mechanism_rl_formulation}, \S\ref{subsec:mechanism_janus}, and \S\ref{subsec:janus_design}}.} in 
Figs.~\ref{fig:coordination_perf}(a) and ~\ref{fig:coordination_perf}(b) on\camii{, respectively,} a performance- and cost-optimized HSS. 
 

We make two key observations.
First, \namePaper{} outperforms its uncoordinated variant by 26.7\% (59.8\%) on average on a performance- (cost-) optimized HSS, as uncoordinated agents make conflicting decisions.
Second, \namePaper{}'s uncoordinated variant performs poorly in read-intensive workloads because the migration agent does not prefetch frequently-read pages to the fast device.

\subsection{\namePaper{}: Design \label{subsec:mechanism_janus}}
\begin{figure}[h]
\centering
\includegraphics[width=\linewidth] {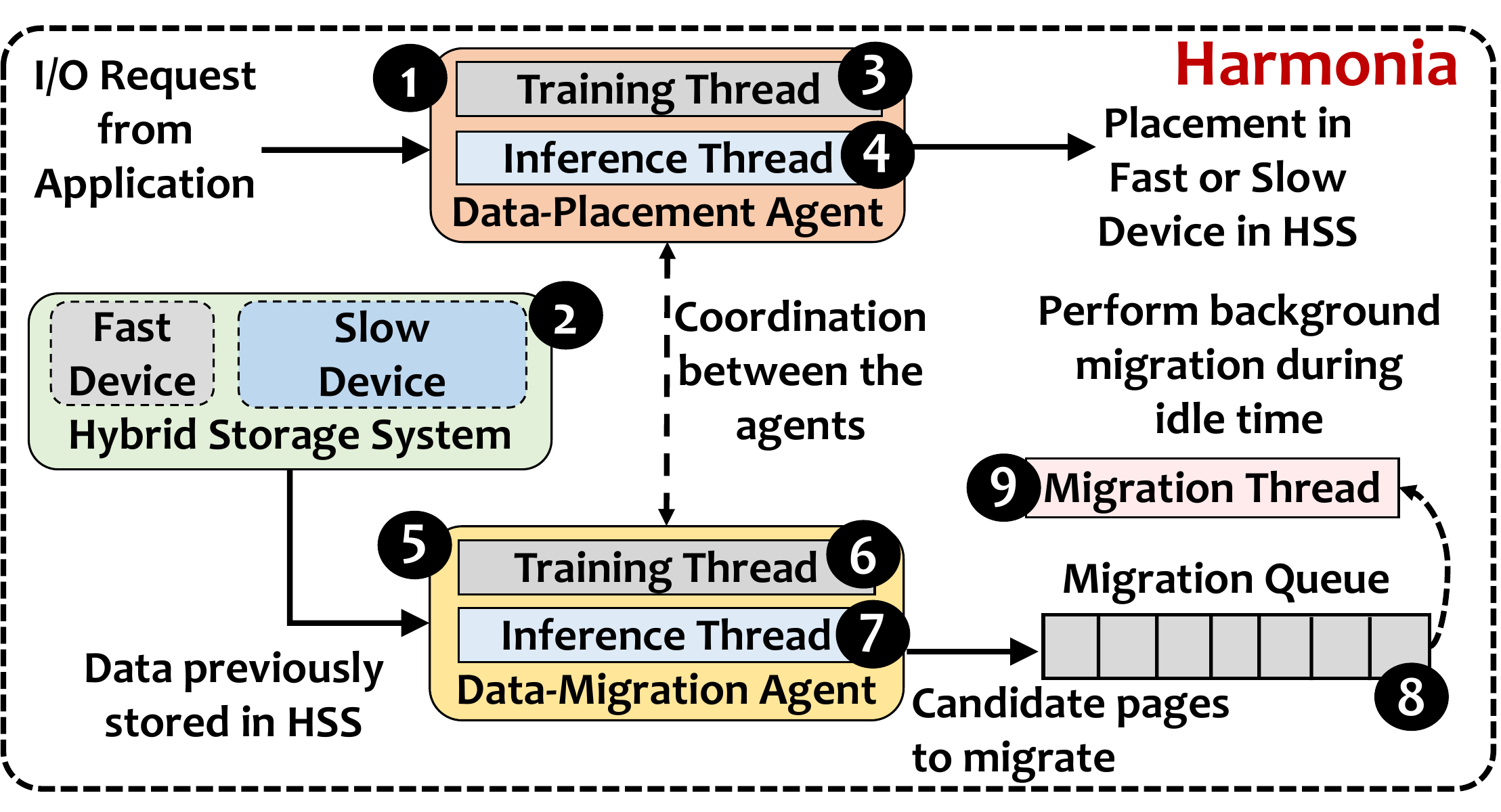}
\vspace{-2.5em}
\caption{Overview of \namePaper{}.}
\label{fig:janus}
\end{figure}

\fig{\ref{fig:janus}} shows \namePaper{}'s key components.

\noindent\textbf{Data-Placement Agent}~\circled{1}. An RL agent that selects the best-fit storage device in the HSS \circled{2} for each incoming I/O write request based on I/O characteristics and HSS conditions.  
It consists of two threads: (1) the \emph{data-placement training thread}~\circled{3}, a background thread which trains the agent and updates the placement policy, and 
(2) the \emph{data-placement inference thread}~\circled{4}, \namePaper{}'s \emph{only} thread, which makes data placement decisions on the \camiv{I/O critical path}.
This separation ensures low-latency data placement decisions while allowing continuous online learning.

\camii{\noindent\textbf{Data-Migration Agent}~\circled{5}.
An RL agent that dynamically reorganizes} previously-placed pages across storage devices by identifying pages for migration and selecting their target devices based on workload access patterns and current HSS conditions.  
To avoid interference with data placement, we implement this agent using two background threads: (1) the \emph{data-migration training thread}~\circled{6}, which trains the agent and continuously updates the migration policy, and (2) the \emph{data-migration inference thread}~\circled{7}, which (i) continuously monitors previously-placed pages,
(ii) determines migration candidates and their target devices, and 
(iii) pushes candidate metadata to the migration queue~\circled{8}.
A page is \emph{not} a migration candidate if its current and target devices are the same. This fully asynchronous design enables proactive data reorganization without increasing data-placement latency on the I/O critical path.
 
\noindent\textbf{Migration Queue} \circled{8}. A queue that stores the metadata of pages chosen for migration by the data migration agent. We store the logical page address and the target device identifier (e.g., 0 for a slow device) of each \camii{page} in the queue.  
Based on empirical analysis (see \S\ref{subsec:perf_results}), we set the queue size to 10 to ensure timely page migration. 

\noindent\textbf{Migration Thread} \circled{9}. A dedicated background process that continuously monitors the migration queue and migrates pages. 
To avoid interference with data placement and enable timely migration, \namePaper{} samples device bandwidth at frequent intervals using real-time monitoring tools (e.g., iostat~\cite{iostat}) to detect device idle periods and schedule migrations.  
For high-intensity workloads (e.g., YCSB~\cite{cooper2010benchmarking}, Baleen~\cite{wong2024baleen, berg2020cachelib, pan2021facebook}), this thread prioritizes pages associated with incoming read requests that are already in the migration queue and issues low-priority writes~\cite{nvme,sata} for migration.

\subsection{Design of \namePaper{}'s RL Agents \label{subsec:janus_design}}
\begin{figure}[h]
\centering 
\includegraphics[width=\linewidth]{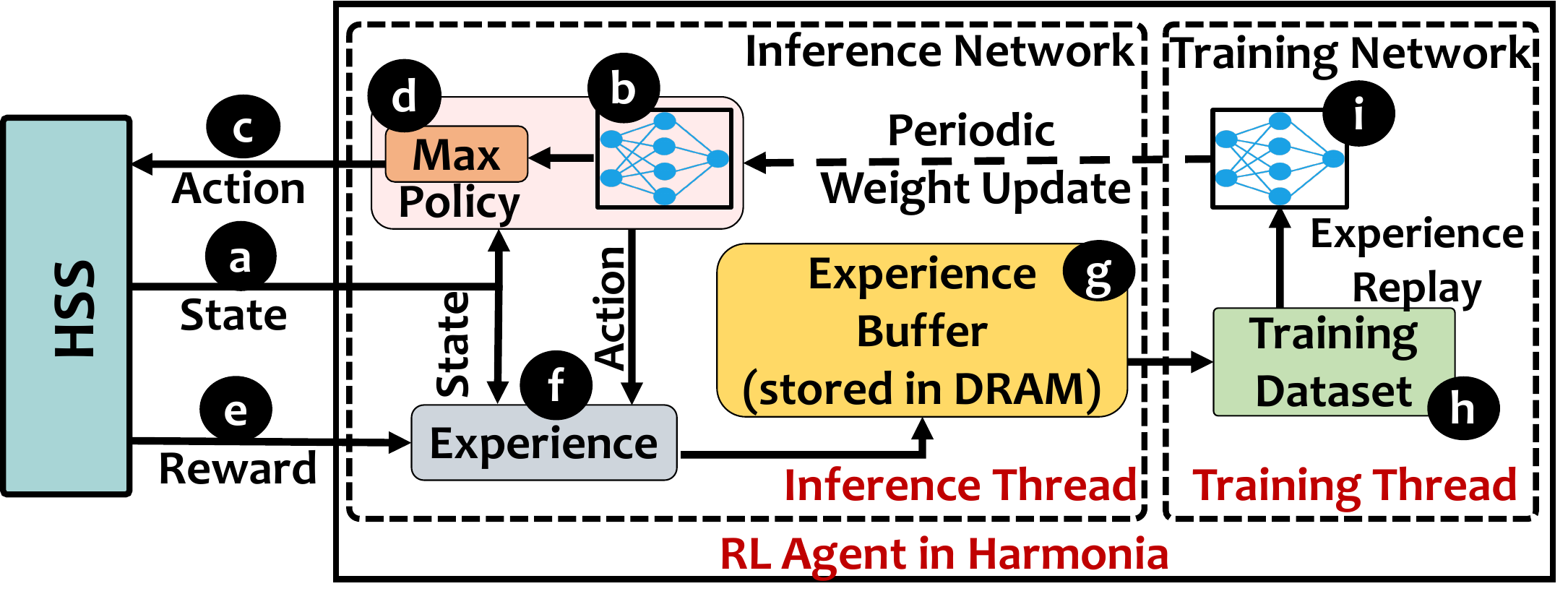}
\vspace{-2.2em}
\caption{Design of \camii{an} RL Agent in \namePaper{}.}
\label{fig:mechanism_rlagent}
\end{figure}

\camii{\fig{\ref{fig:mechanism_rlagent}} shows the design of an RL agent in \namePaper{}}. 
\camii{Each agent uses two threads, a training thread and an inference thread, to enable continuous learning and low-latency inference on the critical path of I/O request handling.} 

\noindent\textbf{Inference Thread}. 
In this thread, \camii{\namePaper{}'s RL agent (1) observes input features (state) \circled{a} from the I/O request and HSS, (2) performs inference using its inference network \circled{b}, (3) selects the action \circled{c} that maximizes \circled{d} \camiv{its} long-term returns, and (4) collects the state, action and reward \circled{e} as experiences \circled{f} in an experience buffer \circled{g}.}

\noindent\textbf{Training Thread}. \camii{In this background thread, \namePaper{}'s RL agent periodically samples collected experiences \circled{h} from the experience buffer to update the training network \circled{i} and optimize the placement or migration policy.}

\head{Experience Buffer}
\camii{Each agent maintains a separate experience buffer to reflect its distinct objective and timescale.} 
\namePaper{} uses experience replay~\cite{dqn,singh2022sibyl} to sample a batch of experiences from \camii{the experience buffer to train each agent}. 
\camii{The experience buffer resides} in the host DRAM and stores the 1000 most recent experiences (based on empirical analysis).
A larger experience buffer captures more access patterns, but increases storage overhead.

\head{Training and Inference Networks}
\camii{Each agent uses} identical simple feed-forward networks~\cite{singh2022sibyl} consisting of an input layer with seven neurons (one per state feature), one hidden layer of 10 neurons~\cite{de1993backpropagation}, and \camii{an output layer with $N$ neurons (where  $N$ is the number of storage devices in the HSS; $N$=2 for dual-HSS)}.
We use the swish activation~\cite{ramachandran2017searching} in the hidden layer, and select the action with the highest predicted Q-value. 
We separate the training and inference networks to remove training latency from I/O request handling. 
\camii{The training network weights are copied to the inference network every $K$ requests to keep the inference network up to date. Based on empirical analysis, we set $K=1000$, which balances timely policy updates and frequent synchronization overhead.} 

\head{RL Algorithm} 
We employ \camii{C51~\cite{bellemare2017distributional}}, a categorical Deep Q-Network~\cite{bellemare2017distributional}, to update the Q-values $Q(S,A)$ (See \S\ref{sec:background}) in \camii{each agent}.
C51 learns the distribution of Q-values for each state-action pair, enabling richer feedback and more stable learning~\cite{harrold2022data,singh2022sibyl}.

\head{Hyper-Parameter Tuning}
\label{subsec:hyperparameters}
We improve \namePaper{}'s accuracy by performing one-time offline hyper-parameter~\cite {paine2020hyperparameter} tuning using cross-validation~\cite{arlot2010survey}. 
Table~\ref{tab:hyperparameter} shows the hyper-parameter values chosen based on empirical analysis.
The discount factor ($\gamma$) balances immediate and future rewards. The learning rate ($\alpha$) controls the rate of neural network weight updates. 
The exploration rate ($\epsilon$) balances exploration and exploitation. 
Batch size determines the number of samples per training iteration.
Experience buffer size is the number of recent experiences stored for training.

\begin{table}[h]
\begin{center}
\caption{\camii{Hyper-parameters of \namePaper{}'s RL Agent}}
\label{tab:hyperparameter}
\vspace{-1em}
\renewcommand{\arraystretch}{1.2}
\setlength{\tabcolsep}{2pt}
\resizebox{\linewidth}{!}{%
\begin{tabular}{|l||c|c|c|}
\hline
\begin{tabular}{c}    
\textbf{Hyper Parameter}
\end{tabular} 
& 
\begin{tabular}{c}    
\textbf{Design Space}
\end{tabular} 
&
\begin{tabular}{c}    
\textbf{Placement} \\ \textbf{Agent}
\end{tabular} & 
\begin{tabular}{c}    
\textbf{Migration} \\ \textbf{Agent}
\end{tabular} \\
\hline
\hline

\begin{tabular}{l}    
\textbf{Discount Factor}
\end{tabular}($\gamma$) & 0-1  & 0.9 & 0.1 \\ \hline
\begin{tabular}{l}    
\textbf{Learning Rate}
\end{tabular} ($\alpha$) & $1e^{-5}-1e^{0}$ & $1e^{-3}$  & $1e^{-2}$ \\ \hline
\begin{tabular}{l}    
\textbf{Exploration Rate}
\end{tabular} ($\epsilon$) & 0-1  & 0.001 & 0.001 \\ \hline
\begin{tabular}{l}    
\textbf{Batch Size}
\end{tabular}  & 64-256   & 128  & 256 \\ \hline
\begin{tabular}{l}    
\textbf{Experience Buffer Size}
\end{tabular} & 10-10000 & 1000 & 1000 \\
\hline
\end{tabular}
}
\end{center}
\end{table}

\head{Exploration vs. Exploitation}
\camii{Each RL agent starts} with no prior knowledge of the workload or HSS, and \camii{makes} \emph{random} initial decisions (\emph{explore}), and \camiv{uses} their experiences (\emph{exploit}) to gradually make optimal decisions.
We use the $\epsilon$-greedy policy~\cite{tokic2011value}, where the predicted best action is selected with (1-$\epsilon$) probability, and a random action with $\epsilon$ probability. 

\begin{figure}[h]
\centering
\includegraphics[width=\linewidth]{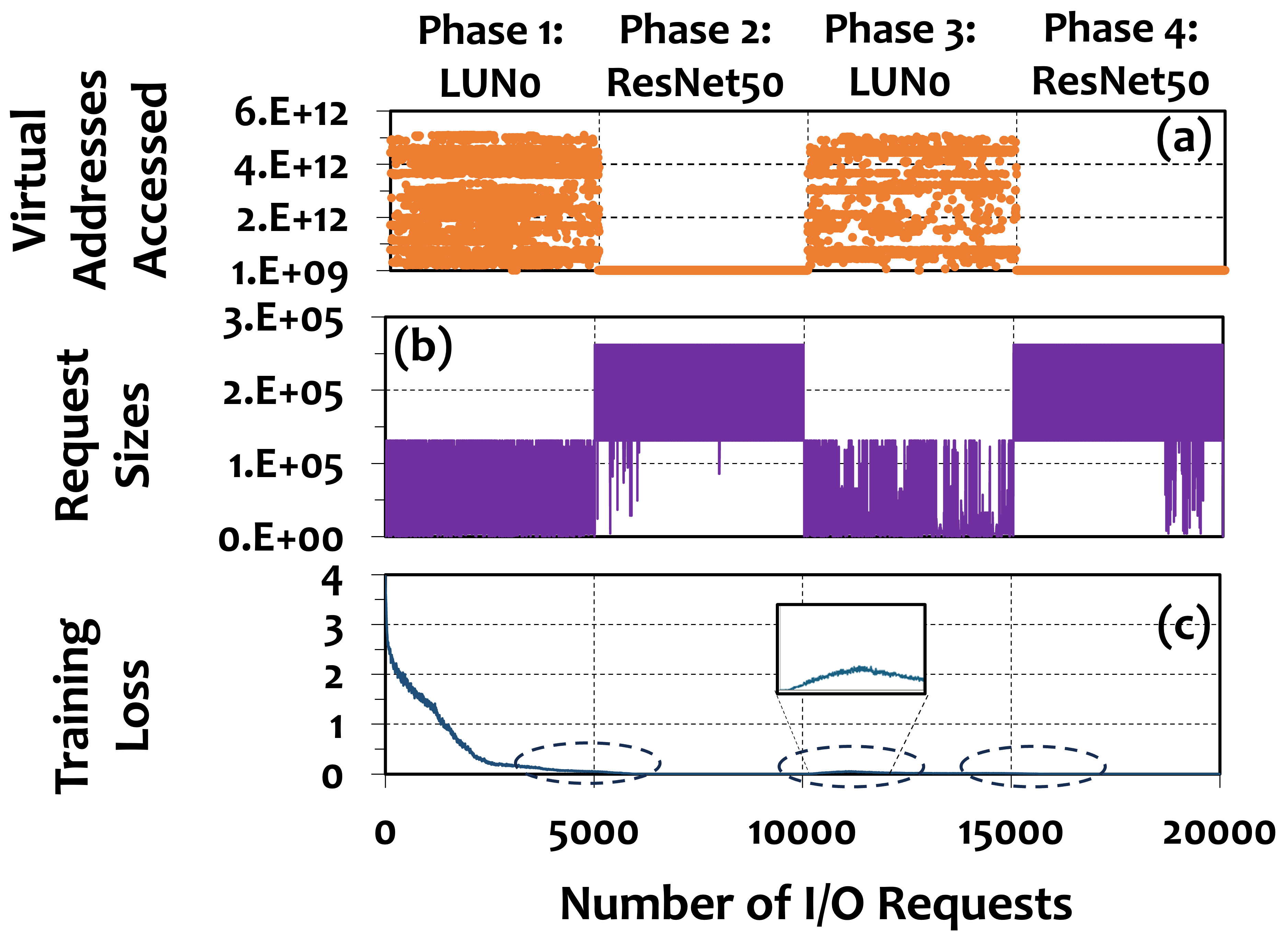}
\vspace{-2em}
\caption{\camii{\namePaper{}'s adaptability to changing workload access patterns. Access patterns of two periodically switching workloads, LUN0 and ResNet50, in terms of (a) virtual addresses accessed, and (b) request sizes. (c) \namePaper{}'s convergence shown as training loss for the first 20000 I/O requests.}}
\label{fig:adaptability}
\end{figure}

\head{\camii{Convergence and Adaptability}}
To evaluate \namePaper{}'s adaptability to changing access patterns, we design a benchmark that \camii{periodically switches} between two workloads, LUN0 (write-intensive) and ResNet50 (read-intensive) (see Table \ref{tab:workloads}),  every 5000 requests. 
\camii{This periodic workload switch creates abrupt changes in access patterns.}
\camiv{Figs.~\ref{fig:adaptability}(a) and~\ref{fig:adaptability}(b)} show the access pattern variations in terms of virtual addresses accessed and request sizes. 
\camiv{Fig.~\ref{fig:adaptability}(c)} shows \namePaper{}'s convergence (in terms of training loss) for 20000 I/O requests on a performance-optimized HSS.
Both agents start with no prior training.

\camii{We make three key observations: 
(1) The policies of both agents converge (i.e., their training loss becomes near zero) within the first 5000 requests (Phase 1: LUN0).   
(2) When the workload changes from LUN0 (Phase 1) to ResNet50 (Phase 2), training loss remains near zero, indicating that previously learned policies generalize well across workload changes.
(3) Small transient spikes (see inset plot) in training loss occur at each phase change. However, the training loss remains low, and \namePaper{}'s policies quickly converge again (within a few hundred requests), demonstrating fast adaptivity to changing access patterns.}

\head{Fallback Mechanism}
\camii{\namePaper{} supports fallback to baseline system behavior to avoid random initial decisions during the exploration phase of each RL agent.
\namePaper{} can be disabled at runtime, reverting placement and migration decisions to those of the default OS policy. 
This allows system administrators to (1) enable \namePaper{} only after each RL agent has learned stable policies, or (2) disable \namePaper{} for workloads with predictable access patterns (e.g., metadata-intensive sequential workloads) where \namePaper{}'s adaptive policies may provide limited benefits.
}

\subsection{Overhead Analysis \label{subsec:overhead_analysis}}

\subsubsection{Storage Overhead. \label{subsubsec: storage_overhead}}
\namePaper{}’s storage overhead consists of five key components: (1) neural networks, (2) experience buffers, (3) migration queue, (4) migration reward, and (5) metadata table.
First, each agent uses a 7–10–2 feed-forward network with 16-bit floating-point weights. 
Across four networks (training + inference for two agents), each with 90 (7$\times$10+10$\times$2) 16-bit weights, the memory footprint is 6 KiB ($\sim$1.5 KiB$\times$4), which fits entirely in CPU caches.
Second, \namePaper{} maintains two experience buffers (one per agent) in DRAM. Each buffer stores 1000 recent experiences and requires 100 KiB, resulting in a total memory footprint of 200 KiB. 
Third, the migration queue stores \camii{a} logical block address (32 bits) and \camii{a} target device ID (4 bits) for 10 migration candidates, requiring 50 B of DRAM.
Fourth, \namePaper{} temporarily stores 50 I/O latencies \camii{(determined based on empirical analysis in \S\ref{subsec:perf_results})} to assign a delayed reward to the migration agent, which consumes \camiv{$\sim$100 B}.
Fifth, \namePaper{} maintains a metadata table~\cite{tsukada2021metadata} that stores the state features. 
\camii{Each entry consumes 32 bits (see Table~\ref{tab:state}), which is 0.1\% of a 4 KiB page size.}

\subsubsection{Latency Overhead} 
\namePaper{} incurs low latency overhead due to its (1) lightweight RL agents, (2) separation of training and inference, and (3) background execution of non-critical threads (e.g., data-migration inference thread).
We perform training and inference on the host CPU, as the neural networks are lightweight and their weights fit in the \camii{CPU's L1 data cache in our evaluated system.}

\head{Inference latency}
\camii{Each inference operation requires 90 MAC operations (7$\times$10+10$\times$2).
On our evaluated system (see Table~\ref{tab:devices}), each inference consumes $\sim$90 CPU cycles (\emph{240 $ns$}) per core, lower than the I/O read latency of a high-end SSD ($\sim$3 $\mu$s)~\cite{inteloptane, samsung2017znand}.}
Only the data-placement agent's inference contributes to I/O latency because it executes on the critical path of I/O request handling.
\camii{The migration agent performs inference in the background and does not lie on the critical path of I/O request handling. Hence, the migration agent's inference latency does not directly affect the data placement of ongoing I/O requests.}

{\noindent{\head{Training latency}}}  
\namePaper{} executes \camii{each RL agent's training} in the background, and \camii{the training} latency does not \camii{directly} impact I/O requests.
Each training step processes 16 batches, and each batch contains 128 training samples.
Each training step involves 184,320 MAC operations \camiv{(i.e., 16 $\times$ (128$\times$7$\times$10+128$\times$10$\times$2))}.
Each training step takes $\sim$200,000 cycles per core ($\sim$53 $\mu$s) on our \camii{evaluated eight-core CPU (see Table~\ref{tab:devices})}.

{\noindent{\head{Metadata Collection Latency}}}
\camii{\namePaper{}'s custom Linux block driver (see \S\ref{sec:evaluation_methodology}) intercepts I/O requests, extracts state features (see \tab{\ref{tab:state}}), and updates the metadata table. 
We measure metadata collection latency using kernel-level timestamps in our implementation.
On our evaluated system (see Table~\ref{tab:devices}), for each I/O request, (1) metadata collection requires $\sim$100 $ns$, and (2) a metadata table update in DRAM consumes 1 $\mu$s.}

\section{Evaluation Methodology\label{sec:evaluation_methodology}}

\begin{figure}[h]
\centering
\includegraphics[width=0.85\linewidth]{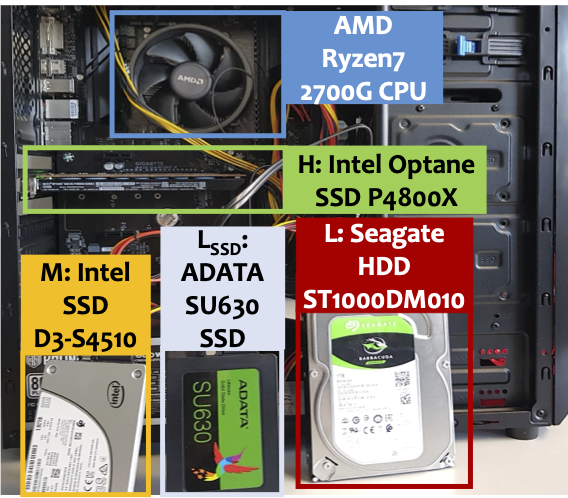}
\vspace{-1em}
\caption{\camiv{Real-system evaluation setup.}}
\label{fig:real_system}
\end{figure}

\camii{We evaluate \namePaper{} on a \emph{real system} (see \fig{}\ref{fig:real_system} and Table \ref{tab:devices}) with four HSS configurations: (i) two dual-HSS (HSS with two storage devices), (ii) tri-HSS (HSS with three storage devices), and (iii) quad-HSS (HSS with four storage devices).}
The dual-HSS configurations include: (1) a performance-optimized HSS, combining a high-end SSD~\cite{inteloptane} with a mid-range SSD~\cite{intels4510}, and (2) a cost-optimized HSS, pairing the high-end SSD~\cite{inteloptane} with a low-end HDD~\cite{seagate}. 
These configurations represent common deployment tradeoffs between performance and storage capacity in modern HPC and datacenter storage nodes.
We implement \namePaper{}'s RL agents using the TF-Agents library~\camii{\cite{TFAgents}}, and design a custom Linux block device driver to interface with the HSS.

\begin{table}[h]
\begin{center}
\caption{{Host System and HSS Configurations}}
\label{tab:devices}
\vspace{-1em}
\renewcommand{\arraystretch}{1.2} %
\setlength{\tabcolsep}{2pt} %
\resizebox{\linewidth}{!}{%
\begin{tabular}{lll}
\hline
\multicolumn{1}{|c|}{\textbf{Host System}}                  & 
\multicolumn{2}{c|}{\begin{tabular}[c]{@{}c@{}}AMD Ryzen 7 2700G~\cite{amdryzen}, \\8-cores@3.5 GHz, 8$\times$64/32 KiB L1-I/D, \\4 MiB L2, 8 MiB L3, \\16 GiB RDIMM DDR4 2666 MHz
\end{tabular}} \\ 
\hline
\hline
\multicolumn{1}{|c|}{\textbf{\camii{Real} Storage Devices}} & \multicolumn{2}{c|}{\textbf{Characteristics}} 
\\ 
\hline
\hline
\multicolumn{1}{|c|}{\textsf{\textbf{H}}: Intel Optane SSD P4800X~\cite{inteloptane}}     & \multicolumn{2}{c|}{\begin{tabular}[c]{@{}c@{}}375 GB, PCIe 3.0 NVMe, SLC, \\ R/W: 2.4/2 GB/s, \\random R/W: 550000/500000 IOPS\end{tabular}} \\ \hline
\multicolumn{1}{|c|}{\textsf{\textbf{M}}: Intel SSD D3-S4510~\cite{intels4510}} & \multicolumn{2}{c|}{\begin{tabular}[c]{@{}c@{}}1.92 TB, SATA TLC (3D), \\ R/W: 560/510 MB/s, \\ random R/W: 895000/21000 IOPS\end{tabular}}
\\ \hline
\multicolumn{1}{|c|}{\textsf{\textbf{L}}: Seagate HDD ST1000DM010~\cite{seagate} } & \multicolumn{2}{c|}{\begin{tabular}[c]{@{}c@{}}1 TB, SATA 6Gb/s 7200 RPM \\ Max. Sustained Transfer Rate: 210 MB/s\end{tabular}}
\\ \hline
\multicolumn{1}{|c|}{\textsf{\textbf{L$_{\mathbf{SSD}}$}}: ADATA SU630 SSD ~\cite{adatasu630} } & \multicolumn{2}{c|}{\begin{tabular}[c]{@{}l@{}}960 GB, SATA, TLC,\\ R/W: 520/450 MB/s\end{tabular}} \\ \hline

\multicolumn{1}{|c|}{\begin{tabular}[c]{@{}c@{}}\textsf{\textbf{PMEM (Emulated)}}: Intel Optane \\ Persistent Memory 200 Series ~\cite{intelpmem200}\end{tabular}} & \multicolumn{2}{c|}{\begin{tabular}[c]{@{}l@{}}128 GB, Memory Mode\\ R/W: 7.45/2.25 GB/s (256B)\end{tabular}} \\ 
\hline
\hline

\multicolumn{1}{|c|}{\textbf{\camii{Real} HSS Configurations}}              & \multicolumn{2}{c|}{\textbf{Devices}} \\ \hline \hline
\multicolumn{1}{|c|}{Performance-Optimized \camii{Dual-HSS}} & \multicolumn{2}{c|}{{high-end (\textsf{H})} \& middle-end (\textsf{M})}\\ \hline 
\multicolumn{1}{|c|}{Cost-Optimized \camii{Dual-HSS}} & 
\multicolumn{2}{c|}{high-end (\textsf{H}) \& low-end (\textsf{L})}\\ \hline
\multicolumn{1}{|c|}{\camii{Dual-HSS} with PMEM} & 
\multicolumn{2}{c|}{Emulated PMEM (\textsf{PMEM}) \& high-end SSD (\textsf{H})}\\ \hline
\multicolumn{1}{|c|}{Tri-HSS} 
& \multicolumn{2}{c|} {\begin{tabular}[c]{@{}c@{}} high-end (\textsf{H}) \& middle-end (\textsf{M}) \& low-end (\textsf{L}) \end{tabular}}
\\\hline
\multicolumn{1}{|c|}{Quad-HSS} & \multicolumn{2}{c|}{\begin{tabular}[c]{@{}c@{}}high-end (\textsf{H}) \& middle-end (\textsf{M}) \& \\ low-end SSD (\textsf{L$_{\mathrm{SSD}}$}) \& low-end HDD (\textsf{L})\end{tabular}}
\\ \hline                              
\end{tabular}
}
\end{center}
\end{table}

\head{Baselines} We compare \namePaper{} against: 
(1) four state-of-the-art heuristic- and machine-learning-based data-placement and migration techniques: \cde{}~\cite{matsui2017design}, \rnn{} (adapted from ~\cite{doudali2019kleio}), \ksvm{}~\cite{shetti2019machine}, and \sibyl{}~\cite{singh2022sibyl}, 
(2) \camii{four extended data-management techniques}, \sibyl{}+\ksvm{}, \sibyl{}+\rnn{}, \cde{}+\rlmigr{} and \sapm{} (single-agent RL formulation for data placement and migration), and   
(3) two ideal approaches: Fast-Only, 
and \oracle{}~\cite{meswani2015heterogeneous} (See \S~\ref{sec:motivation_limitations} for detailed description of the baselines).
Fast-Only assumes that the fast device is large enough to hold the entire workload.
Oracle makes optimal decisions based on knowledge of future access patterns and serves as a reference to evaluate \namePaper{}'s decision accuracy.

\head{Workloads}
We select 25 data-intensive storage workloads (see Table~\ref{table:trace}) from \camii{SYSTOR17}~\cite{lee2017understanding}, RocksDB traces~\cite{yadgar2021ssd}, YCSB~\cite{cooper2010benchmarking}, MLPerf Storage\footnote{We generated the traces by running MLPerf applications, LLM inference and training on our system.}~\cite{mlperfstoragebenchmarks}, large language model (LLM) inference and training~\cite{huggingface, paszke2019pytorch, touvron2023llama, zhang2022opt}, and Baleen~\cite{wong2024baleen, berg2020cachelib, pan2021facebook} from real enterprise, cloud, and datacenter environments.

\begin{table}[h]
    \centering
    \setlength{\tabcolsep}{2pt}
    \scriptsize
    \caption{Characteristics of the Evaluated Workloads}
    \label{tab:workloads}
    \vspace{-1.5em}
    \renewcommand{\arraystretch}{1.3}
    \resizebox{\linewidth}{!}{
    	\begin{tabular}{|c||c|c|c|c|}\hline
    	\begin{tabular}{c}\textbf{Benchmark} \\ \textbf{Suite} \end{tabular} &	\textbf{Traces} & \textbf{Read} \% &
            \begin{tabular}{c}    
            \textbf{Avg. Request} \\\textbf{Size (KB)}
            \end{tabular}
             &
             \begin{tabular}{c}    
            \textbf{Avg. Inter-} \\\textbf{Request Time ($\mu$s)}
            \end{tabular}
              \\
                \hline
                \hline
                                
                \multirow{6}{*}{\begin{tabular}{c}\textbf{SYSTOR17} \\~\cite{lee2017understanding}
                \end{tabular}}
                & LUN0 & 0.2 & 31.7 & 1163.9\\\cline{2-5}
                & LUN1 & 0.3 & 34.2 & 1864.1\\\cline{2-5}
    		& LUN2 & 7.6 & 31.1 &       
                1418.9\\\cline{2-5}
                & LUN3 & 3.5 & 42.7 & 734.5\\\cline{2-5}
                & LUN4 & 0.5 & 26.3 & 823.1\\\hline
                
                \multirow{6}{*}{\begin{tabular}{c}\textbf{RocksDB}\\~\cite{yadgar2021ssd}\end{tabular}}
                & ssd-00 & 79.9 & 108.9 & 66.4\\\cline{2-5}
    		& ssd-01 & 73.5 & 75.1 & 
                40.7\\\cline{2-5}
                & ssd-02 & 79.9 & 7.5 & 3.3\\\cline{2-5}
                & ssd-03 & 79.9 & 9.5 & 3.5\\\cline{2-5} 
                & ssd-04 & 79.9 & 7.8 & 3.6\\\hline

                \multirow{6}{*}{\begin{tabular}{c}\textbf{YCSB}\\~\cite{cooper2010benchmarking}\end{tabular}}
                & YCSB-B & 51.3 & 45.9 & 9.3\\\cline{2-5}
                & YCSB-C & 47.6 & 54.6 & 6.5\\\cline{2-5}
                & YCSB-D & 55.9 & 36.1 & 8.5\\\cline{2-5}
                & YCSB-E & 52.1 & 46.6 & 9.6\\\cline{2-5}
    		& YCSB-F & 49.5 & 53.1 & 6.6\\\hline

                \multirow{2}{*}{\begin{tabular}{c}\textbf{MLPerf Storage}\\~\cite{mlperfstoragebenchmarks}\end{tabular}}
                & ResNet50 & 80.0 & 172.6 & 500.1\\\cline{2-5}
    		& CosmoFlow & 83.4 & 180.1 & 1023.8\\\hline

            \multirow{3}{*}{\begin{tabular}{c}\textbf{LLM}\\~\cite{huggingface, paszke2019pytorch, touvron2023llama, zhang2022opt}\end{tabular}}
                & LLM-Checkpoint & 10.5 & 256 & 94.8\\\cline{2-5}
                & LLM-Weights & 25.1 & 128.0 & 45.0\\\cline{2-5}
                & LLM-KVCache & 45.0 & 16.0 & 24.0\\\hline

            \multirow{5}{*}{\begin{tabular}{c}\textbf{Baleen}~\cite{wong2024baleen, berg2020cachelib, pan2021facebook}\\ (Unseen Workloads)\end{tabular}}
                & Region 1 & 87.9 & 3578.9 & 7.1\\\cline{2-5}
                & Region 2 & 78.1 & 2781.6 & 6.5\\\cline{2-5}
                & Region 3 & 74.4 & 2084.7 & 7.6\\\cline{2-5}
                & Region 4 & 79.9 & 2850.8 & 4.0\\\cline{2-5}
                & Region 5 & 66.8 & 2920.6 & 3.0\\\hline
    	\end{tabular} 
	}
	\label{table:trace}
\end{table}

Each workload has a data footprint of \textbf{$\sim$50000$\times$} the fast device capacity, which ensures sustained capacity pressure on the HSS.
We choose these workloads because they (1) exhibit diverse and complex I/O behavior, including varying read/write ratios, request sizes, spatial/temporal locality, and inter-request arrival times, and (2) stress storage devices with write amplification, read disturbance, and spatial locality effects arising from compaction and log-structured updates.\camii{\footnote{\camii{We provide a detailed workload characterization in the extended version of this paper~\cite{nadig2025harmonia}.}}}
Each workload runs in a separate thread. 

\camii{SYSTOR17 workloads~\cite{lee2017understanding}} capture storage traffic from enterprise virtual desktop infrastructure (VDI) deployment. 
\camii{RocksDB~\cite{yadgar2021ssd}} represents key-value store workloads, which affect write amplification, and spatial locality. 
\camii{YCSB~\cite{cooper2010benchmarking}} captures cloud database operations (e.g., insert, update, read, scan).
\camii{MLPerf Storage~\cite{mlperfstoragebenchmarks}} captures modern machine learning and data analytics pipelines, such as image classification and parameter prediction.
We include three LLM workloads capturing the dominant I/O behaviors during training and inference. 
\camii{LLM-Checkpoint (e.g., ~\cite{touvron2023llama, maurya2024datastates, liu2026adacheck})} captures periodic checkpoint writes and recovery reads during 7B-parameter LLM (e.g., LLaMA-2-7B~\cite{touvron2023llama}) training, which generates bursty sequential writes followed by read-dominated recovery phases.
\camii{LLM-Weights (e.g., ~\cite{zhang2022opt, song2024powerinfer})} captures repeated loading of model parameters for inference using a 7B-parameter LLM (e.g., OPT-6.7B~\cite{zhang2022opt}), which stresses cold-start latency, large sequential reads, and page-cache interactions. 
\camii{LLM-KVCache (e.g., ~\cite{wu2026crosskv, jang2026cost, jiang2025efficient, wang2026multi})} records fine-grained KV-cache spill and reload behavior, which produces small, latency-sensitive reads and writes with rapidly shifting locality.
We include Baleen workloads~\cite{wong2024baleen, berg2020cachelib, pan2021facebook}, which \camii{capture} production ML data \camii{pipelines} and model-serving I/O behavior from Meta's production environment. We do \emph{not} use Baleen workloads for hyper-parameter tuning\camii{; they} serve as unseen workloads to evaluate \namePaper{}'s adaptability.

\head{Multi-Programmed Workloads} To represent real-world scenarios, we generate six \emph{multi-programmed}  workloads by running multiple workloads concurrently (see Table~\ref{tab:mixed_workloads}).
We construct these workloads based on three key factors: (1) a combination of read-intensive, write-intensive, and mixed workloads, (2) the number of concurrent workloads required for high I/O intensity, and (3) different request sizes.   
Each constituent workload runs on a single thread (e.g., mix6 runs on eight concurrent threads).
Multi-programmed workloads typically exhibit higher I/O request intensity and varied access patterns.

\begin{table}[h]
\caption{Characteristics of multi-programmed workloads}
\label{tab:mixed_workloads}
\vspace{-1em}
\centering
\setlength{\tabcolsep}{2pt}
\resizebox{\linewidth}{!}{%
\begin{tabular}{|c|c|l|}
\hline
\textbf{Mix} & 
\begin{tabular}{c}    
\textbf{Constituent} \\\textbf{Workloads~\cite{cooper2010benchmarking, lee2017understanding, yadgar2021ssd}}
\end{tabular} 
&
\makecell[c]{\textbf{Description}} \\
\hline
\hline
\textbf{mix1} &  
ssd-02 and LUN4 & 
\begin{tabular}{l}
ssd-02 is read-intensive \\ and LUN4 is write-intensive
\end{tabular}
\\
\hline
\textbf{mix2}                      & \begin{tabular}{l}
LUN1 and ssd-04
\end{tabular} &
\begin{tabular}{l}
\camiv{LUN1} is write-intensive \\ and ssd-04 is read-intensive
\end{tabular}
\\
\hline
\textbf{mix3}                     & 
\begin{tabular}{c}
YCSB-C and YCSB-F
\end{tabular}
&
\begin{tabular}{l}
Both have near-equal read-write ratio
\end{tabular}
\\
\hline
\textbf{mix4}                     & 
\begin{tabular}{c}
ssd-00, ssd-04, \\\camiv{LUN0 and LUN1}
\end{tabular}
&
\begin{tabular}{l}
Two read-intensive \\ and two write-intensive workloads
\end{tabular}
\\
\hline
\textbf{mix5}                     & 
\begin{tabular}{c}
ssd-00, LUN0, \\YCSB-C and YCSB-F
\end{tabular}
&
\begin{tabular}{l}
Read-intensive, write-intensive and \\ 
two workloads with a \\ near-equal read-write ratio
\end{tabular}
\\ 
\hline
\textbf{mix6}                     & 
\begin{tabular}{c}
YCSB-B, YCSB-D, LUN0, LUN1, \\
LUN4, ssd-00, ssd-02, \camiv{ssd-03}
\end{tabular}
&
\begin{tabular}{l}
Two with near-equal read-write ratio,\\
three write-intensive \\ and three read-intensive
\end{tabular}
\\
\hline
\end{tabular}
}
\end{table}

\camii{\head{Metrics}
In our evaluation (see \S\ref{sec:results}), we report four key metrics to evaluate \namePaper{} and prior approaches. 
First, \emph{average I/O request latency} is the average time required to complete an application I/O request. 
Second, \emph{end-to-end throughput}, measured in IOPS (i.e., number of I/O operations per second), captures the overall request-processing rate of the HSS. 
Third, \emph{tail latency}, measured as the 99th and 99.99th percentile of I/O request latency, characterizes the slowest response time observed by the applications.
Fourth, \emph{write amplification (WA)} (e.g.,~\cite{hu2009write, sun2013measuring, lu2013extending, niu2018hybrid, micheloni2018hybrid}), is the ratio of the total data written to all the HSS storage devices (including migration-induced writes) to the data writes originating from the workload. A WA value of 1 indicates that the HSS performs no additional writes beyond the application-issued writes; i.e., the total data written to the HSS storage devices equals the data written by the workload. A higher WA adversely impacts device \camii{lifetime}.
}
\section{Evaluation}\label{sec:results}
\subsection{Performance Analysis \label{subsec:perf_results}}
\begin{figure}[h] 
\centering
\includegraphics[width=\linewidth]{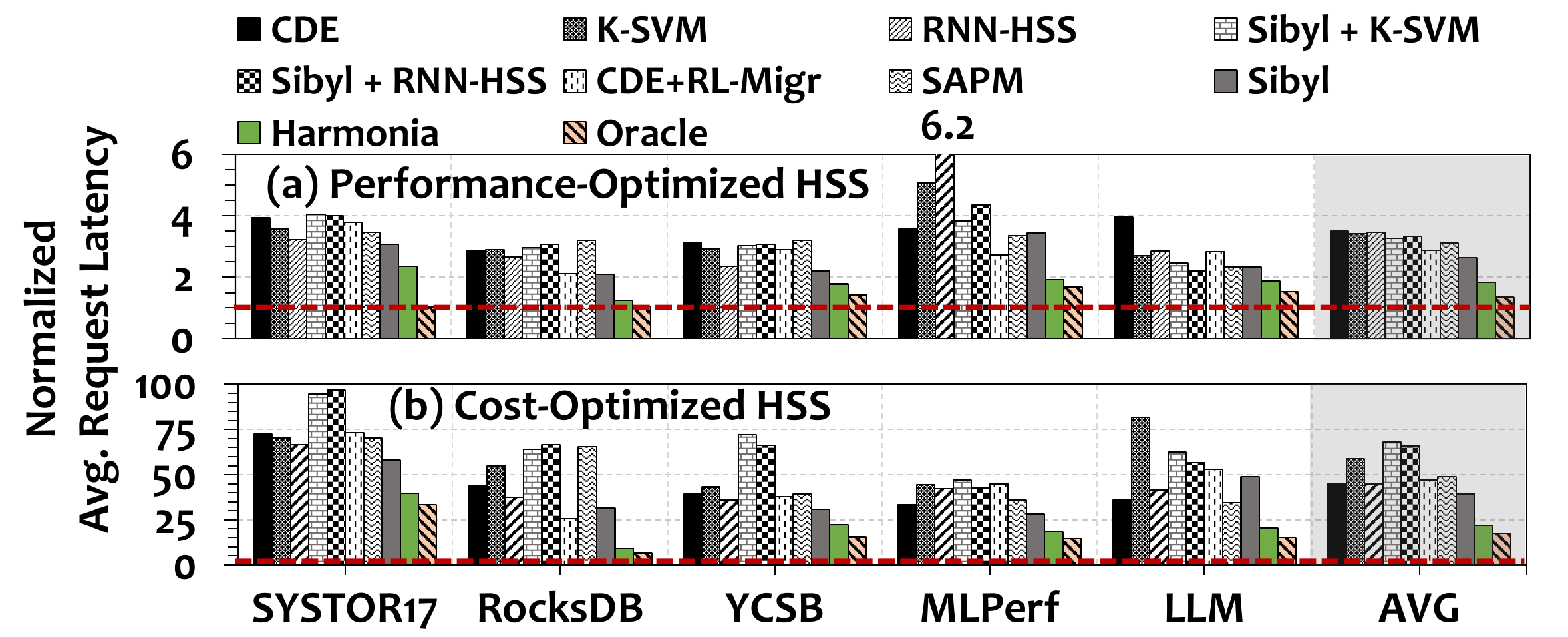}
\vspace{-2em}
\caption{\camii{Average I/O request latency of \namePaper{} and baselines on (a) performance-optimized and (b) cost-optimized HSS, normalized to the latency of Fast-Only}. Lower is better.}
\label{fig:eval_perf}
\end{figure}

\begin{figure}[h]
\centering
\includegraphics[width=\linewidth]{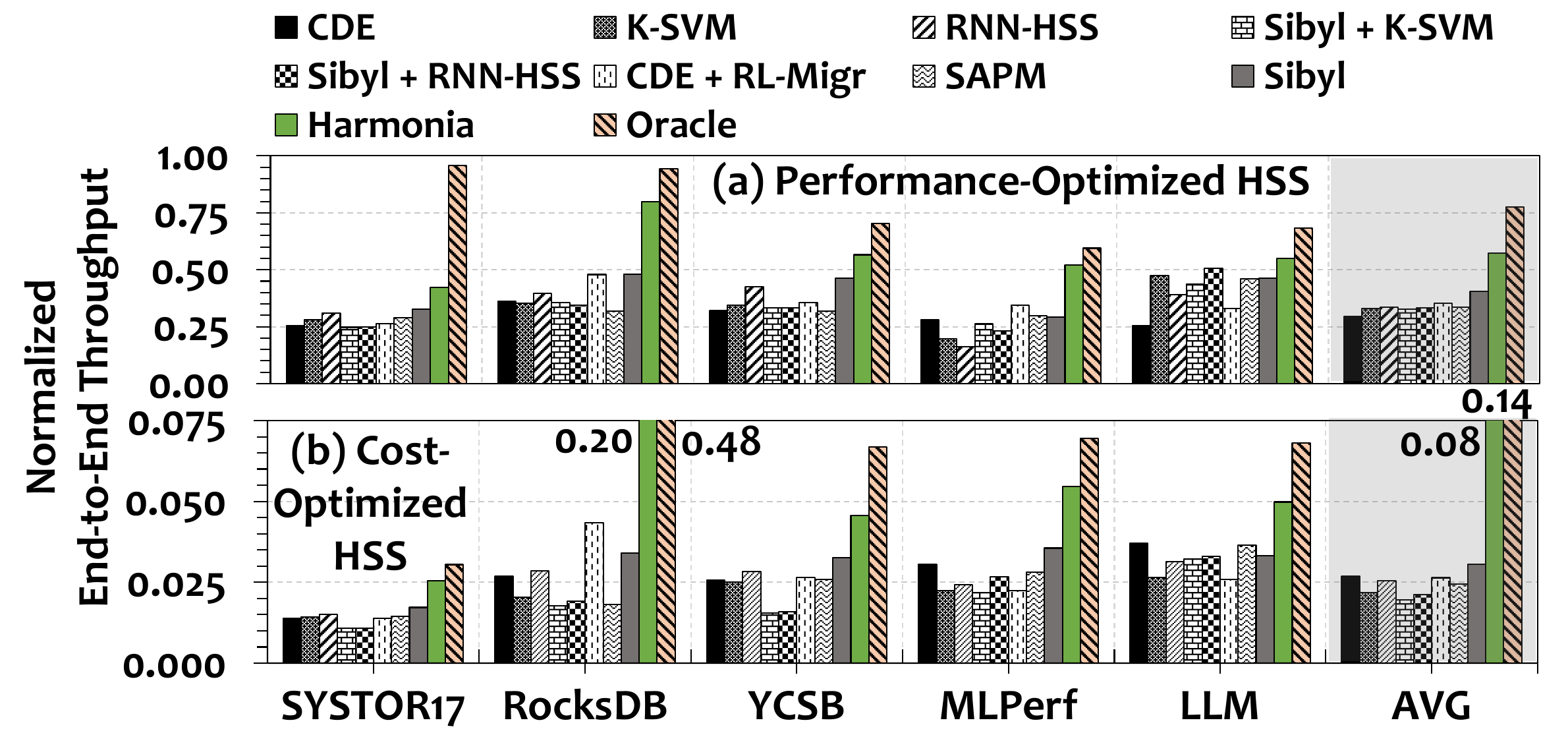}
\vspace{-2em}
\caption{\camii{End-to-end throughput (in IOPS) of \namePaper{} and baselines on (a) performance-optimized and (b) cost-optimized HSS, normalized to the throughput of Fast-Only.} Higher is better.}
\label{fig:eval_throughput}
\end{figure}

\head{I/O Request Latency and End-to-End Throughput} 
\figs{\ref{fig:eval_perf}(a)} and ~\ref{fig:eval_perf}(b) show the \camii{average I/O request latency} of \cde{}, \ksvm{}, \rnn{}, \sibyl{}, \sibyl{}+\ksvm{}, \sibyl{}+\rnn{}, \cde{}+\rlmigr{}, \sapm{}, \namePaper{} and \oracle{} on performance- and cost-optimized HSS.
\figs{\ref{fig:eval_throughput}(a)} and \ref{fig:eval_throughput}(b) show their end-to-end throughput \camii{(see metrics in \S\ref{sec:evaluation_methodology})}. 
\camii{The average I/O request latency and end-to-end throughput of \namePaper{} and baselines are normalized to the corresponding values} of Fast-Only. We make four key observations.

First, \namePaper{} consistently outperforms all baselines \camiii{on} both HSS configurations. 
On a performance- (cost-) optimized HSS, \namePaper{} improves \camii{latency} by 29.3\% (44.8\%) and end-to-end throughput by 49.4\% (136.7\%) over the best-performing prior approach, \sibyl{}, on average. These gains demonstrate the benefit of jointly optimizing both placement and migration. 

Second, \namePaper{} \camii{provides} performance close to that of an \oracle{}. On average, \camii{\namePaper{}'s latency is 41.0\% (33.1\%) higher than \oracle{}'s latency, and provides 75.5\% (69.2\%) of \oracle{}'s throughput on a performance- (cost-) optimized HSS}. This result indicates that \namePaper{}'s coordinated online learning policy makes \camii{effective} decisions without knowledge of the workload's future access patterns.

Third, \namePaper{}'s \camii{latency} and throughput improvements over baselines are higher in read-intensive workloads (e.g., RocksDB, MLPerf) than in write-intensive workloads (e.g., SYSTOR17). 
For example, on average, \namePaper{} outperforms \sibyl{} by 57.5\% (43.2\%) in read-intensive workloads and 41.0\% (39.9\%) in write-intensive workloads on a \camiv{performance- (cost-)} optimized HSS. 
In read-intensive workloads, unlike \sibyl{}, \namePaper{} proactively prefetches frequently-accessed pages to the fast device even without application writes, reducing read latency and improving throughput. 
In write-intensive workloads, \camii{compared to \sibyl{},} \namePaper{} further improves \camii{latency} by proactively migrating cold pages from the fast device to accommodate new incoming writes. 
In MLPerf and LLM workloads, \namePaper{} learns to keep \camiii{frequently-reused} training data on the fast device and migrates cold data in the background, which reduces fast-storage contention compared to prior approaches.

Fourth, \namePaper{}'s \camii{latency} and throughput improvements over baselines are higher on a cost-optimized HSS than on a performance-optimized HSS because: 
(1) migration overheads are larger on a cost-optimized HSS than on a performance-optimized HSS due to the large latency gap between the two devices, 
(2) several baselines perform migration on the critical path of I/O request handling, which adds significant latency overheads, and 
(3) \camii{\namePaper{} proactively migrates cold pages from the fast device during system idle times, which frees space in the fast device without interfering with incoming I/O requests.}

\begin{figure}[h]
\centering
\includegraphics[width=\linewidth]{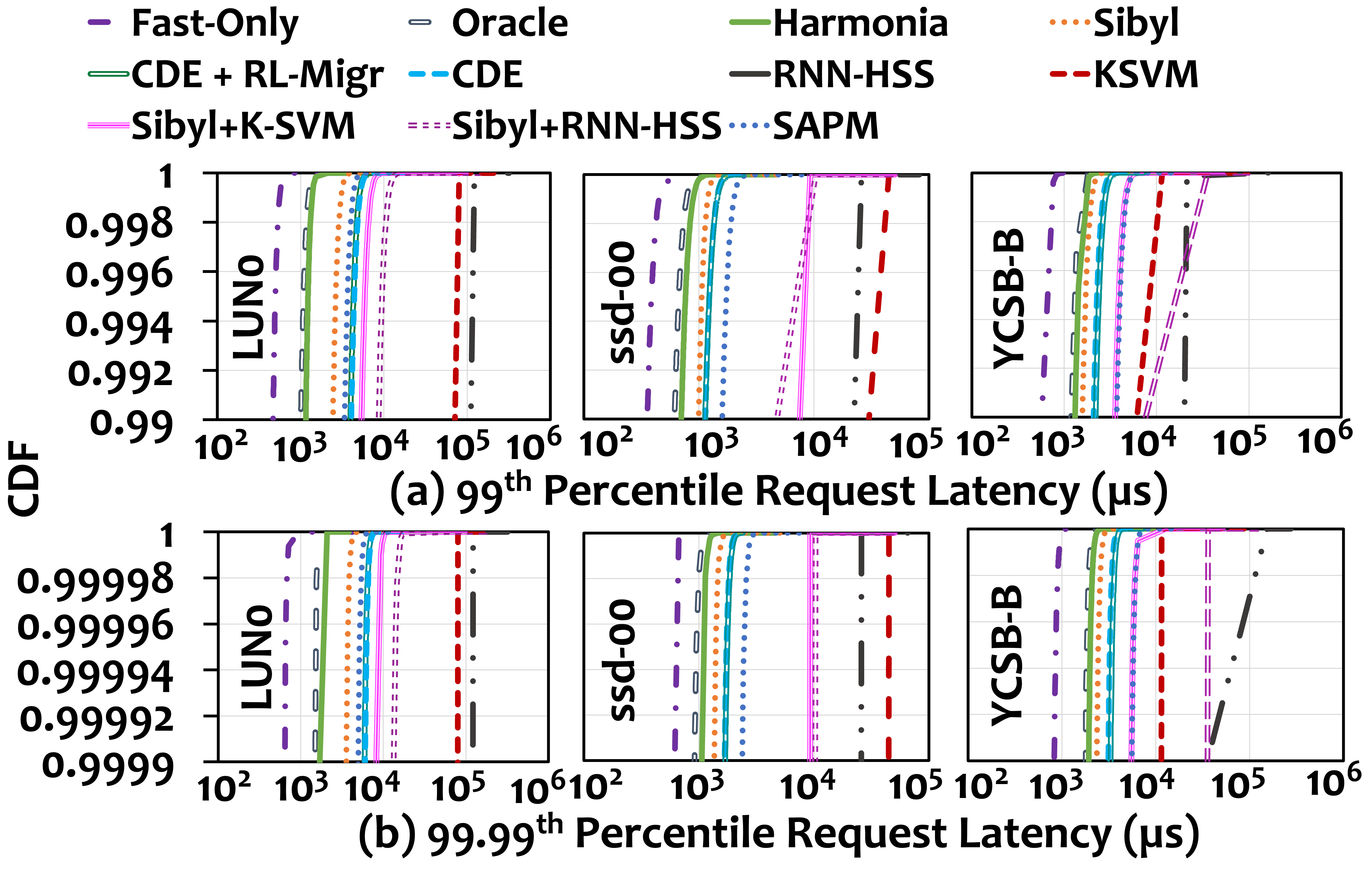}
\vspace{-2em}
\caption{99th (top) and 99.99th (bottom) percentile tail latencies of \namePaper{} and baselines on three workloads, LUN0, ssd-00 and YCSB-B, on a performance-optimized HSS.}
\label{fig:tail_latency}
\end{figure}

\head{Tail Latency}  
\figs{\ref{fig:tail_latency}(a)} and \ref{fig:tail_latency}(b) show the 99th and 99.99th percentile I/O latencies (\emph{tail latency}) of \namePaper{} and baselines on a performance-optimized HSS using three representative workloads: LUN0 (write-intensive), ssd-00 (read-intensive) and YCSB-B (mixed) (see \S\ref{sec:evaluation_methodology}).
We make two key observations.

First, \namePaper{} significantly reduces tail latency compared to all baselines across the three workloads. 
On average, it lowers the 99th (99.99th) percentile I/O latency by 32.3\% (25.3\%) compared to \sibyl{}. These gains come from performing data migration during low-utilization periods, avoiding interference with latency-critical I/O requests on the critical path.
Second, \ksvm{}, \rnn{}, \sibyl{}+\ksvm{} and 
\sibyl{}+\rnn{} show very high tail latencies because they perform bulk migrations at fixed intervals. These bulk migrations introduce device contention and queueing delays, which adversely affect tail latency.

\begin{figure}[h]
\centering
\includegraphics[width=\linewidth] {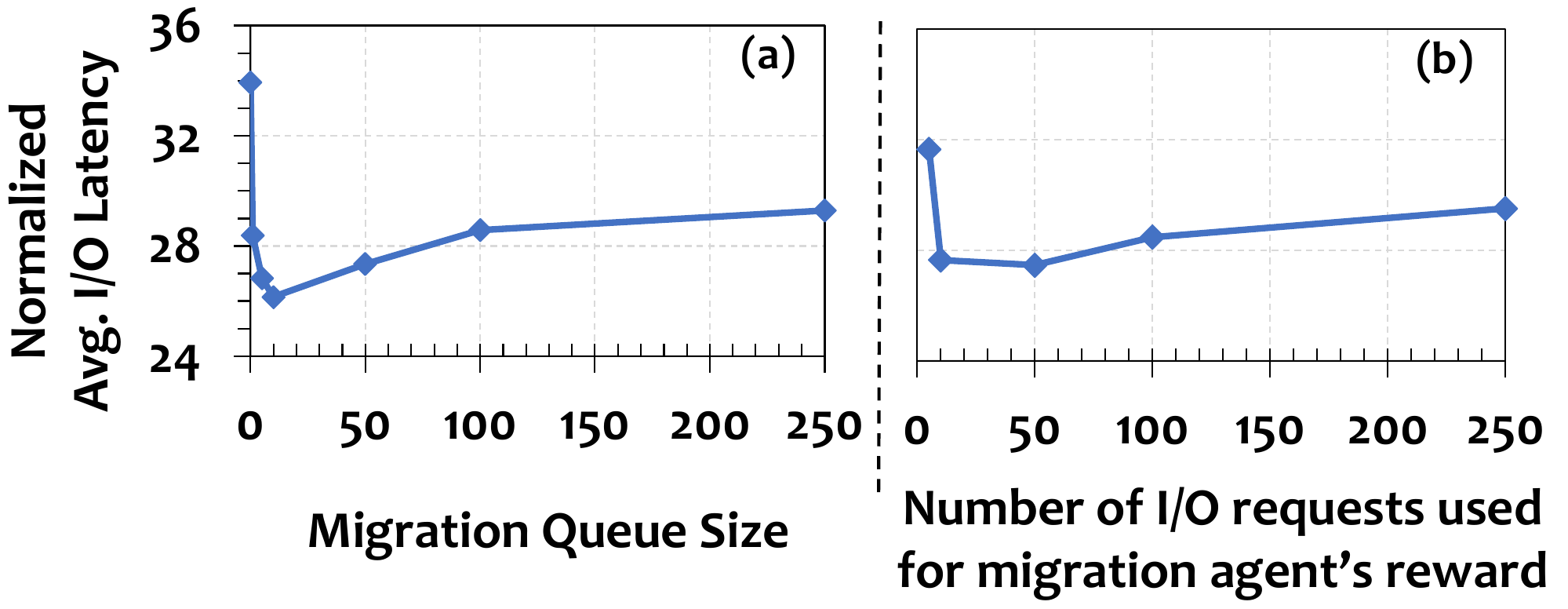}
\vspace{-2em}
\caption{\camii{Latency} sensitivity to (a) migration queue sizes, and (b) number of future incoming requests used in migration agent's reward structure.}
\label{fig:sensitivity}
\end{figure}

\head{Performance Sensitivity to Migration Queue Size}
\fig~\ref{fig:sensitivity}(a) shows the impact of migration queue size on \namePaper{}'s \camii{average I/O request latency}. We make three key observations.
First, a migration queue size of 10 entries provides the \camii{lowest latency} in our evaluated configurations.
Second, low queue \camii{sizes} (i.e., < 5) limit the number of migration candidates, which results in performance degradation due to dropping many useful migration candidates. 
If the queue size is 0, \namePaper{} performs no data migration and exhibits poor performance.
Third, large queues store more migration candidates, but increase the likelihood of older entries becoming stale. This causes migration decisions not to reflect current system conditions.

\head{Performance Sensitivity to Migration Agent's Reward}
\fig~\ref{fig:sensitivity}(b) shows the \camii{latency impact} of the number of future I/O requests ($n$) used to compute the migration agent's delayed reward (See Equation \ref{eq_Janus_migration_reward}). 
\camii{The value of $n$ determines how far into the future the migration agent evaluates the impact of its decisions, and its optimal value depends on several factors, including (1) inter-request times of incoming I/O requests after migration, (2) workload access patterns (e.g., access frequency and reuse distance), and (3) system conditions (e.g., device capacities).}
We make three key observations. 

First, the \camii{best value of $n$ in our evaluated configuration is 50}. 
\camii{Second, small values of $n$ (e.g., < 10) do not capture the benefits of migration because the impact of migration decisions is visible \emph{only} after a sufficient number of future I/O requests occur. Two factors contribute to this behavior: (i) when migration frees space on the fast device, we observe latency improvement \emph{only} when there are many subsequent placements that utilize the freed capacity in the fast device, and (ii) when migration prefetches pages to the fast device, we observe latency improvement \emph{only} after there are many accesses to those pages.
} 
\camii{Third, large values of $n$ (e.g., 250) degrade performance because the impact of migration reward is diluted if we consider: (i) a large number of future I/O requests that may be unrelated to the previously-migrated pages, 
or (ii) I/O requests that arrive too far into the future, which are \emph{unlikely} to benefit from the freed fast-device capacity.}
 
\begin{figure}[h]
\centering
\includegraphics[width=\linewidth] {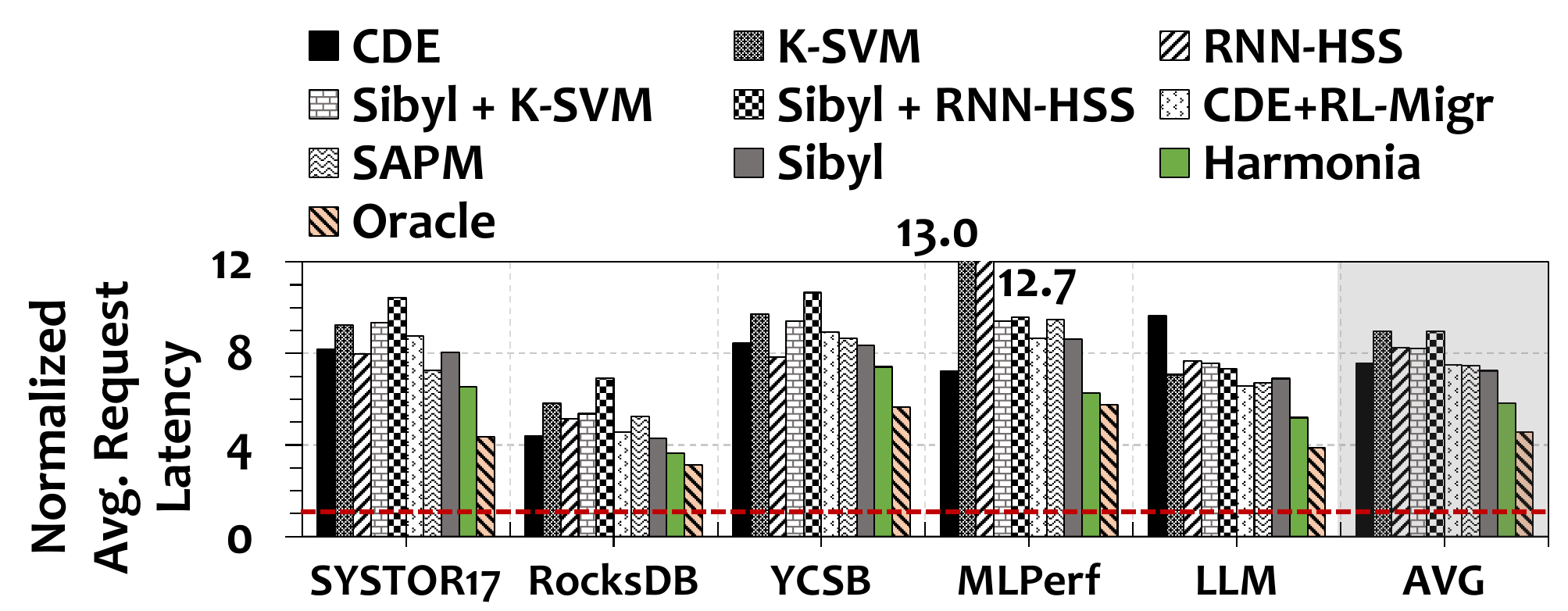}
\vspace{-2em}
\caption{\camii{Average request latency} of \namePaper{} and baselines on an HSS with PMEM~\cite{intelpmem200} and high-end SSD~\cite{inteloptane}, normalized to that of Fast-Only. Lower is better.}
\label{fig:pmem_performance}
\end{figure}

\head{Performance on HSS with Persistent Memory (PMEM)} 
To demonstrate extensibility to emerging memory technologies, we evaluate \namePaper{} and baselines on an HSS with a persistent memory (PMEM)~\cite{intelpmem200} and a high-end SSD~\cite{inteloptane} (see Table \ref{tab:devices}). 
We emulate PMEM using a RAMDisk configured to match PMEM's latency characteristics, as these emerging devices are not readily available.
\fig{\ref{fig:pmem_performance}} shows the \camii{average I/O request latency} of \namePaper{} and baselines on HSS with PMEM normalized to Fast-Only (i.e., PMEM stores all data). We make three key observations.

First, on average, \namePaper{} outperforms the best-performing baseline, \sibyl{}, by 19.4\%. \camii{\namePaper{} has \emph{only} 28.0\% higher latency than the \oracle{}.}
These benefits come from \namePaper{}'s joint optimization of both placement and migration policies.
\camii{Second, \cde{} and \cde{}+\rlmigr{} place all incoming data in the PMEM device, and exploit PMEM's low write latency.
However, PMEM devices typically provide much lower capacity than SSDs due to their higher cost per bit and lower storage density. Hence, PMEM's limited capacity leads to frequent page migrations that consume bandwidth and interfere with ongoing I/O requests. These frequent migrations diminish the benefits of aggressive placement on the PMEM device. 
}
Third, \namePaper{} provides \camii{larger latency} improvements on read-intensive workloads (i.e., RocksDB, MLPerf) because it proactively prefetches data to the PMEM device, resulting in more reads from PMEM.    
We discuss \namePaper{}'s applicability to emerging memory technologies in \S\ref{sec:discussion}.

\begin{figure}[h]
\centering
\includegraphics[width=\linewidth]{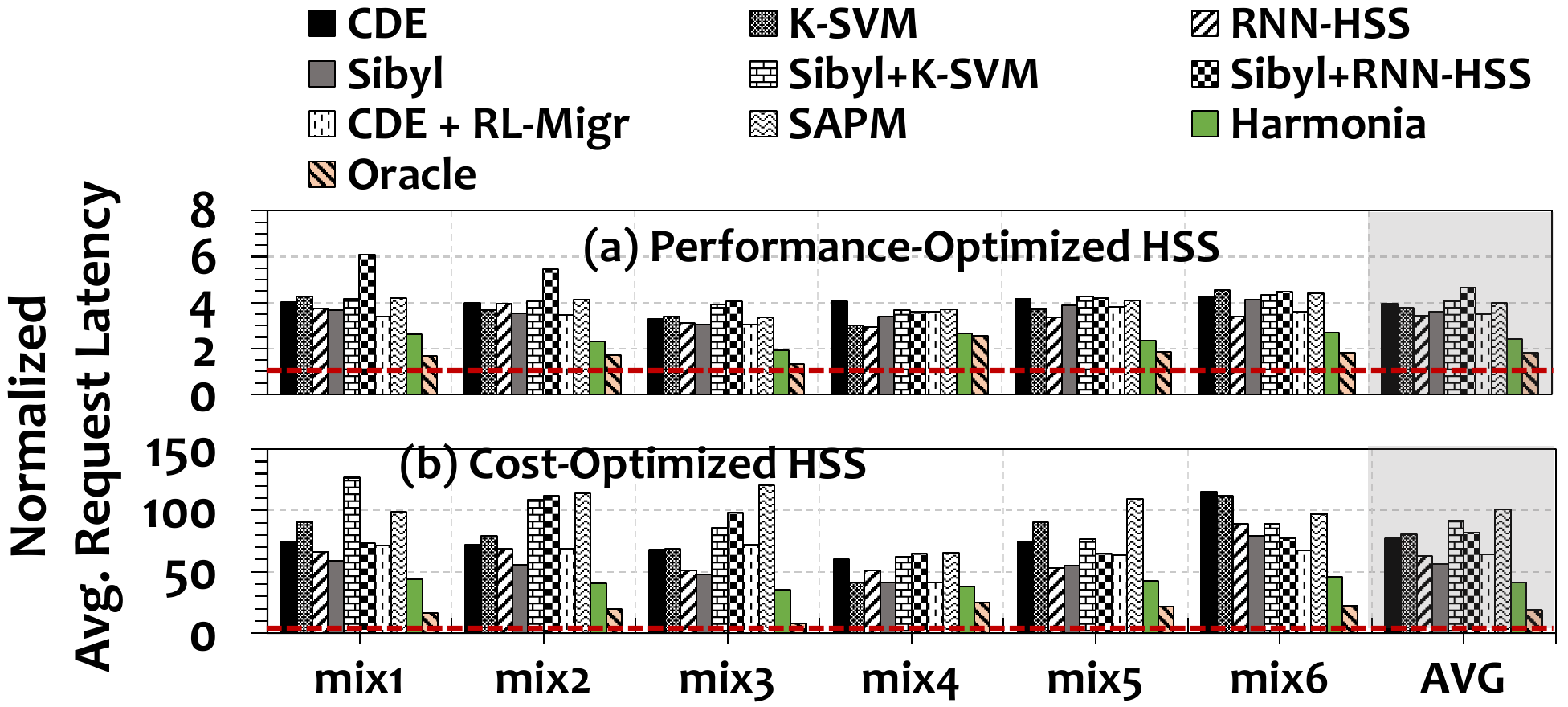}
\vspace{-2em}
\caption{\camii{Average request latency} of \namePaper{} and baselines on multi-programmed workloads on (a) performance-optimized, and (b) cost-optimized HSS. Latency values are normalized to those of Fast-Only. Lower is better.}
\label{fig:eval_mpworkloads}
\end{figure}

\head{Performance on Multi-Programmed Workloads} 
\figs{\ref{fig:eval_mpworkloads}}(a) and \ref{fig:eval_mpworkloads}(b) show the \camii{average I/O request latency} of \namePaper{} and baselines on multi-programmed workloads (see Table \ref{tab:mixed_workloads}) on performance- and cost-optimized HSS. 
We make two key observations.

First, \namePaper{} outperforms prior approaches in all workloads and HSS configurations. 
On a performance- (cost-) optimized HSS, \namePaper{} (i) \camii{outperforms} \sibyl{} by 32.8\% (25.3\%), and \camii{(ii) has 35.4\% (54.7\%) higher latency than that of an \oracle{}}.  
Second, \namePaper{}'s latency improves with the number of \camii{concurrently-executing} workloads. 
On eight \camii{concurrently-executing} workloads (i.e., mix6), \namePaper{} \camii{provides} the highest \camii{latency improvements}, 39.4\% (41.8\%), over \sibyl{}. 
As concurrency increases, workload access patterns become more dynamic, and interference between applications increases.
\namePaper{}'s RL-based policies adapt more quickly to access pattern variations than prior approaches.

\begin{figure}[h]
\centering
\includegraphics[width=\linewidth]{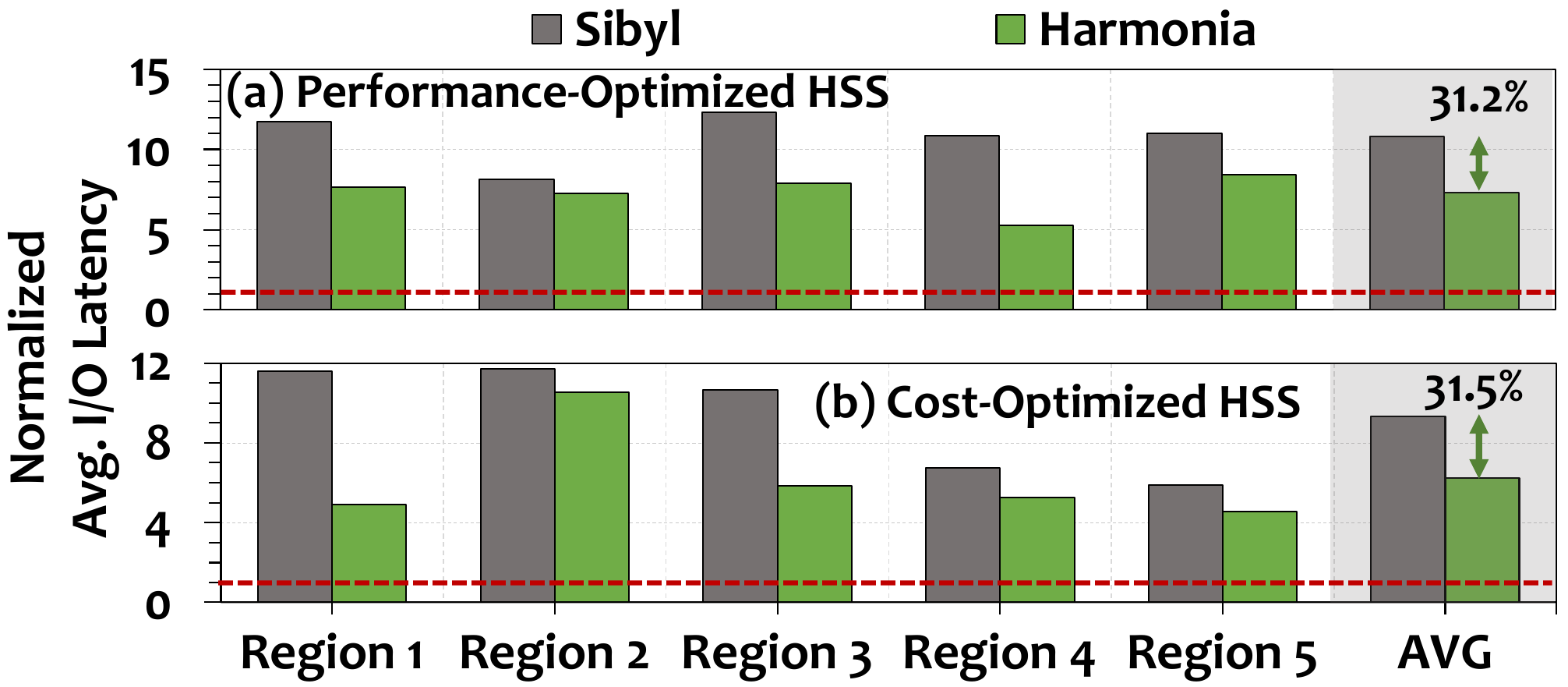}
\vspace{-2em}
\caption{\camii{Average request latency of \namePaper{} and \sibyl{} on unseen workloads on (a) performance- optimized, and (b) cost-optimized HSS, normalized to the latency of Fast-Only}. Lower is better.}
\label{fig:eval_unseen}
\end{figure}

\head{Performance on Unseen Workloads}
We evaluate \namePaper{}'s \camii{average I/O request latency} on five unseen workloads~\cite{wong2024baleen, berg2020cachelib, pan2021facebook} from Meta's production environment, which are not used during hyperparameter tuning (see \S\ref{sec:evaluation_methodology}). 
\figs{}~\ref{fig:eval_unseen}(a) and \ref{fig:eval_unseen}(b) compare \namePaper{} to \sibyl{} on a performance- (cost-) optimized HSS.
We make two key observations. 

\camiv{First,} \namePaper{} outperforms \sibyl{} by 31.2\% (31.5\%) on average on a performance- (cost-) optimized HSS. These improvements demonstrate that \namePaper{}'s multi-agent RL framework generalizes beyond the workloads used for hyperparameter tuning. 
Second, although \namePaper{}'s hyperparameters are tuned specifically for these workloads, its online-learning-based data-placement and data-migration policies adapt to changing workload access patterns. In contrast, \sibyl{} relies on a heuristic migration policy that \camii{has limited adaptability to} workload changes. 

\subsection{Extensibility to Multiple Devices in HSS}\label{subsec:eval_extensibility}
\begin{figure}[h]
\centering
\includegraphics[width=\linewidth]{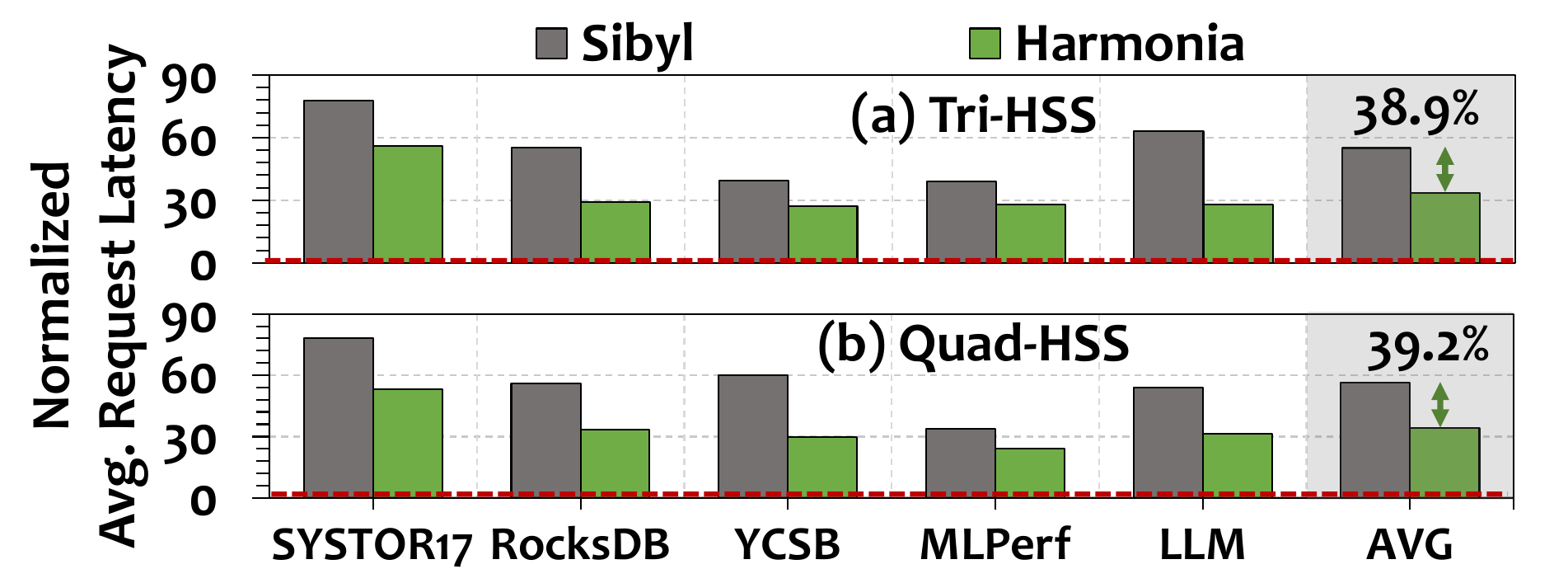}
\vspace{-2em}
\caption{\camii{Average request latency} of \sibyl{} and \namePaper{} on a Tri-HSS and a Quad-HSS, normalized to that of Fast-Only. Lower is better.}
\label{fig:extensibility}
\end{figure}

We evaluate \namePaper{}'s extensibility to different HSS configurations by comparing its performance with \sibyl{} on an HSS with three \camii{devices} (Tri-HSS) and four devices (Quad-HSS) (see Table \ref{tab:devices}).
We do not evaluate other baselines as they are designed for fixed HSS configurations and lack easy extensibility~\camii{\cite{singh2022sibyl}}. 
\figs{\ref{fig:extensibility}(a)} and \ref{fig:extensibility}(b) show the \camii{average I/O request latency} of \sibyl{} and \namePaper{} on Tri- and Quad-HSS. 

We make two key observations.
First, \namePaper{} consistently outperforms \sibyl{} on both HSS configurations, improving \camii{latency} by 38.9\% (39.2\%) on average in a Tri-HSS (Quad-HSS). These gains come from placing performance-critical data in high-end and mid-range devices and migrating cold data to low-end devices during idle periods. 
In contrast, \sibyl{} frequently evicts data to the low-end devices on the critical path, introducing significant latency overhead.
Second, in read-intensive workloads (e.g., RocksDB, MLPerf), \namePaper{} \camii{provides} higher \camii{latency improvements} (up to 58.3\%) because \camii{\sibyl{} primarily depends on application writes to migrate data.} 
We \camii{further} discuss \namePaper{}'s extensibility in \S\ref{sec:discussion}.

\subsection{Impact on Device Lifetimes} \label{subsec:eval_wa}
\begin{figure}[h]
\centering
\includegraphics[width=\linewidth]{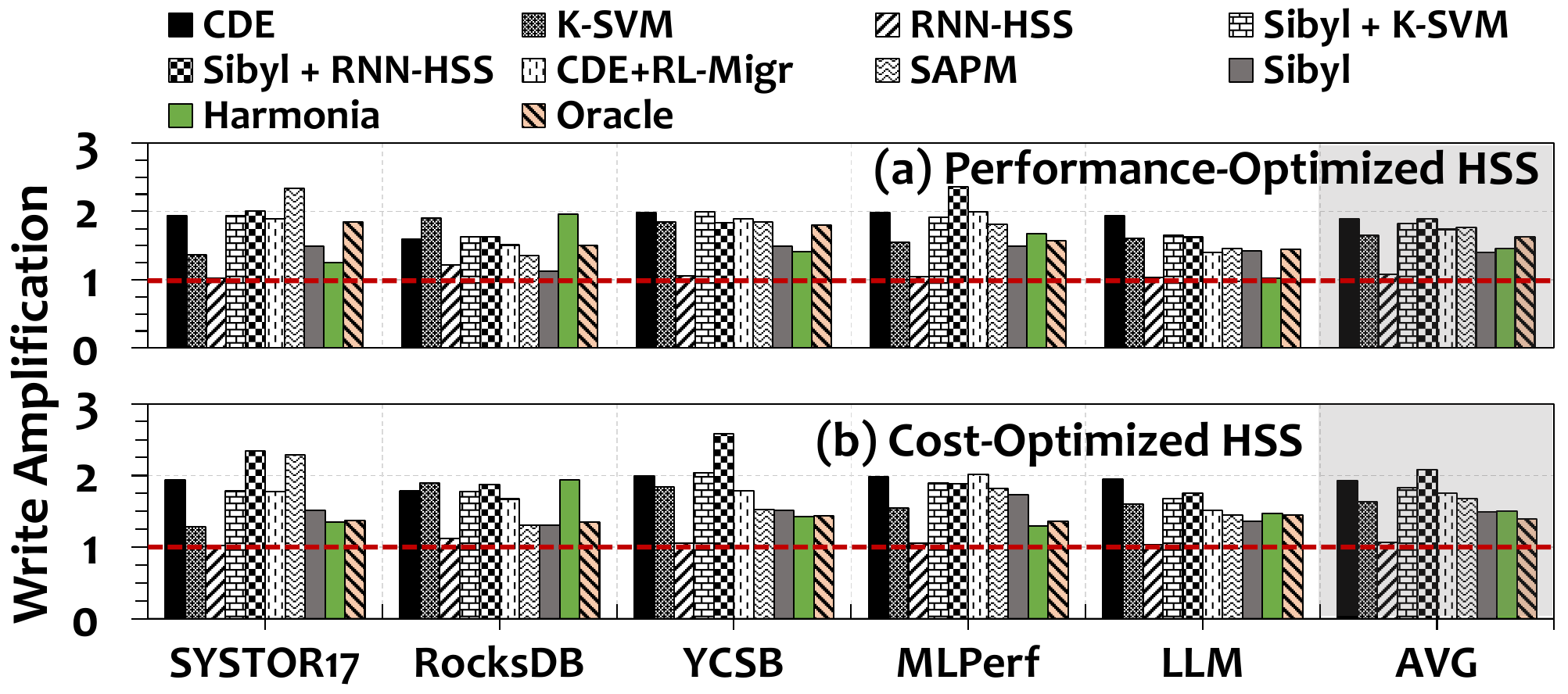}
\vspace{-2em}
\caption{Write amplification of \namePaper{} and baselines on a performance- (cost-) optimized HSS. Lower is better.}
\label{fig:eval_waf}
\end{figure}

We evaluate the impact of \namePaper{} and baselines on storage device lifetimes by measuring their write amplification (WA) \camii{(see metrics in \S\ref{sec:evaluation_methodology}).}
\figs{\ref{fig:eval_waf}(a)} and \ref{fig:eval_waf}(b) show WA of \namePaper{} and baselines on performance- and cost-optimized HSS, respectively.

We make two key observations.
First, only \sibyl{} and \rnn{} have a lower WA than \namePaper{}. 
\namePaper{}'s average WA of 1.44 (1.54) is higher than \sibyl{}'s 1.38 (1.50) and \rnn{}'s 1.09 (1.07) in a performance- (cost-) optimized HSS. \namePaper{}'s higher WA is due to proactive prefetching of frequently accessed data to the fast device in read-intensive workloads to improve read performance. 
Second, \namePaper{}'s WA varies significantly across workload types. In write-intensive workloads (i.e., \camii{SYSTOR17}), \namePaper{} has a lower WA (1.24) compared to its WA (1.8\camii{0}) in read-intensive workloads (i.e., RocksDB, MLPerf) because \namePaper{} opportunistically migrates data during application-triggered updates in write-intensive workloads, reducing the need for additional migration traffic.

We conclude that Harmonia slightly increases write amplification to achieve significant \camii{latency improvements}.

\begin{figure}[h]
\centering
\includegraphics[width=\linewidth]{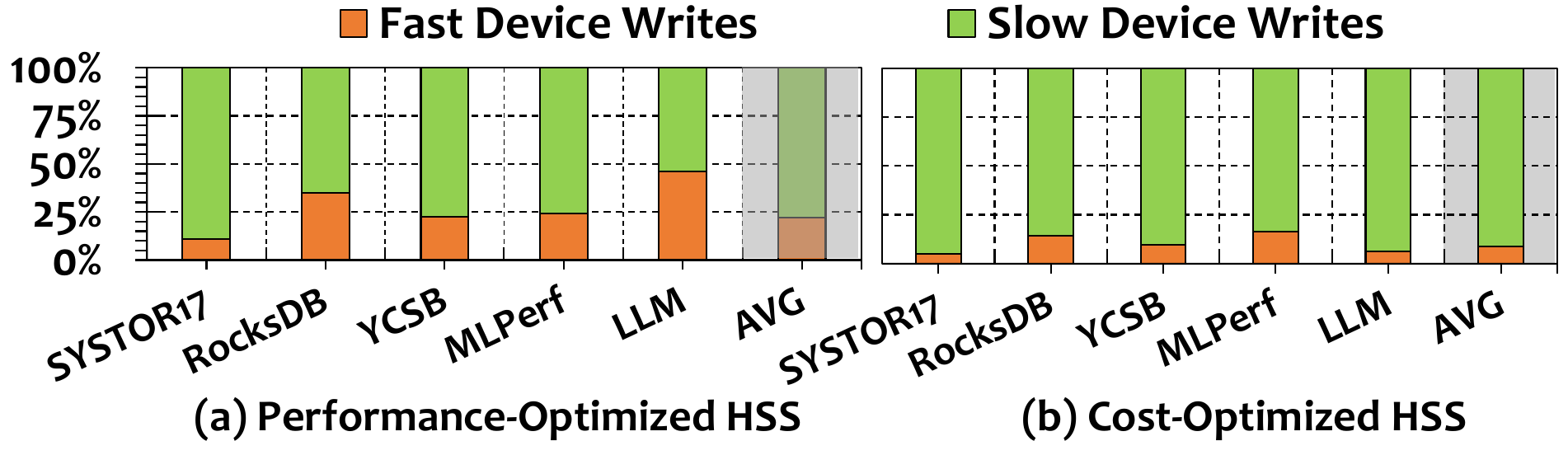}
\vspace{-2em}
\caption{Write traffic due to \namePaper{}'s migrations on performance-optimized (left) and cost-optimized (right) HSS.}
\label{fig:eval_writetraffic}
\end{figure}

\head{Write Traffic Distribution in \namePaper{}}
\fig{}~\ref{fig:eval_writetraffic} shows the distribution of \namePaper{}'s migration-induced writes across devices on performance- and cost-optimized HSS.

We make two observations.
First, \namePaper{} migrates more data to the slow device to preserve the fast-device capacity for frequently-accessed data. 
Second, \namePaper{} performs more writes to the fast device in read-intensive workloads (i.e., RocksDB, MLPerf) because it proactively prefetches frequently-read data to the fast device even when there are no application updates. 
\section{Discussion}
\label{sec:discussion}

\head{Alternate Objectives and Reward Structures}
\namePaper{} currently improves HSS performance using I/O latency-based reward signals, but its RL formulation can be adapted for other objectives such as device lifetime, fairness, quality of service (QoS), and energy efficiency. 
For example, lifetime optimization can incorporate write counts, device endurance limits, write amplification, and device reliability characteristics (e.g., NAND type, ECC) into the state representation and reward structure. 
Fairness and QoS can incorporate metrics such as bandwidth, queue depth, and application requirements. 
These extensions require changes to state representation and reward formulation without modifying \namePaper{}'s architecture. We leave the exploration of these objectives to future work.

\head{\namePaper{}'s Extensibility}
\namePaper{} is easily extensible to HSS configurations with \camii{more than four storage devices} with minimal designer effort. 
Extending \namePaper{} to \camii{support} each additional storage device requires two changes to each agent's training and inference networks (see \S\ref{subsec:janus_design}): (1) adding one additional input node (i.e., state feature) representing the capacity utilization of the new device (see \tab{\ref{tab:state}}), and (2) adding one additional output node representing the new placement or migration action.
\camii{For each additional storage device in the HSS, these changes result in eight additional nodes (i.e., two RL agents $\times$ two neural networks $\times$ two nodes).}
\camii{Together, these eight nodes increase (1) storage overhead by 1.3 KiB, and (2) inference and training latency by $\sim$40 $ns$, which is negligible compared to storage-device access latency.
Each additional device increases the metadata storage overhead by three bits per metadata-table entry (see \tab{\ref{tab:state}}) to store the new device's capacity utilization.}

\head{Applicability to Other Systems}
\camii{\namePaper{}'s design is applicable to other systems such as tiered \camii{memory/storage} (e.g.,~\cite{aguilera2007improving, cherubini2017data, raghavan2014tiera,zhang2010adaptive, zhang2010automated, vengerov2008reinforcement, yang2017autotiering, salkhordeh2015operating, montgomery2014extent, elnably2012efficient, heuristics_usenix_2014, zhang2025rethinking}), hybrid memory (e.g., ~\cite{li2017utility,agarwal2015page_HMM,agarwal2017thermostat_HMM,goglin2016exposing,ham2013disintegrated,lin2016memif,malladi2016dramscale,pavlovic2013data,pena2014toward,qureshi2009scalable,yoon2012row,meza2013case, meza2012enabling, ren2015thynvm, yu2017banshee}), and disaggregated \camii{memory/}storage (e.g., ~\cite{maruf2023tpp, gouk2023memory, li2023pond}).}
In tiered \camii{memory/storage} systems (e.g.,~\cite{aguilera2007improving, cherubini2017data, raghavan2014tiera,zhang2010adaptive, zhang2010automated, vengerov2008reinforcement, yang2017autotiering, salkhordeh2015operating, montgomery2014extent, elnably2012efficient, heuristics_usenix_2014, zhang2025rethinking}), the data-placement policy is often static \camii{(i.e., default placement in the fastest device present in the top tier)}, while the data-migration policy moves data between adjacent tiers. 
For hybrid memory systems \camii{(e.g., ~\cite{li2017utility,agarwal2015page_HMM,agarwal2017thermostat_HMM,goglin2016exposing,ham2013disintegrated,lin2016memif,malladi2016dramscale,pavlovic2013data,pena2014toward,qureshi2009scalable,yoon2012row,meza2013case, meza2012enabling, ren2015thynvm, yu2017banshee})}, we can adapt \namePaper{}'s RL formulation to operate under stricter latency constraints by using lightweight state representations and table-based RL approaches (e.g., ~\cite{bera2021pythia, bera2026athena, ipek2008self, rummery1994line, sutton1998reinforcement}). 
In CXL-based disaggregated \camii{memory/storage} systems (e.g., ~\cite{maruf2023tpp, gouk2023memory, li2023pond}), \namePaper{} can be deployed \camii{in} the CXL fabric controller to coordinate placement and migration decisions across pooled \camii{memory/storage} resources. 
We hope that \namePaper{} can aid in the design of adaptive data-management policies for these architectures.
\section{Related Work \label{sec:related_work}}
\camii{To our knowledge, \namePaper{} is the first multi-agent RL-based holistic data-management technique that jointly optimizes data placement and data migration in hybrid storage systems.
We already qualitatively and quantitatively compared \namePaper{} against state-of-the-art data placement~\cite{singh2022sibyl, matsui2017design} and data migration techniques~\cite{shetti2019machine, doudali2019kleio}. 
We review related works on HSS data management, RL in storage systems, RL for architectural optimizations, and multi-agent RL systems.}

\head{HSS Data Management} A large body of prior work proposes heuristic- and machine-learning-based data-placement (e.g.,~\cite{matsui2017design,sun2013high,heuristics_hyrbid_hystor_sc_2011, lv2013probabilistic, guerra2011cost, elnably2012efficient, heuristics_usenix_2014, bu2012optimization, krish2016efficient, tai2015sla, zhang2010automated, wu2012data, iliadis2015exaplan, lv2013hotness, matsui2017tri, feng2014hdstore, yang2017autotiering, singh2022sibyl, doudali2019kleio,ren2019archivist,cheng2019optimizing,shetti2019machine, li2017utility,agarwal2015page_HMM,agarwal2017thermostat_HMM,ham2013disintegrated, lin2011hot,  salkhordeh2015operating, pavlovic2013data, meza2013case, chou2017batman, kim2011hybridstore, wang2019panthera, ramos2011page, liu2019hierarchical,luo2020optimal,doudali2021cori,sen2019machine, xing2025proactive, zeng2025charten, yan2025io, zhou2024augur, chang2024idt}) and data-migration (e.g.,~\cite{lin2011hot, lin2017efficient, lin2018buffer, zhang2010adaptive, cheng2015amc, shetti2019machine, vengerov2008reinforcement,vasilakis2020hybrid2,yi2025artmem}) techniques.

Linux kernel utilities such \camiv{as} \camii{\emph{mdadm}} and \camii{\emph{logical volume manager (LVM)}} support software RAID and virtual partitions, but leave placement and migration to applications or administrators. 
In contrast, \namePaper{} offers adaptive, application-transparent data placement and migration.
For data migration, Vengerov~\cite{vengerov2008reinforcement} applies fuzzy rules in hierarchical storage.
ArtMem~\cite{yi2025artmem} proposes RL-based page migration for tiered memory. 
\camii{Neither work jointly optimizes page placement and migration. Both works have limited scalability because they rely on a table-based RL approach that maintains explicit state-action tables whose size grows rapidly with the number of storage devices and state features.}

\head{RL in Storage Systems}
Several prior works (e.g., \cite{liu2019learning,yoo2020reinforcement,wang2020reinforcement, kang2018dynamic, li2023rlalloc, li2020mitigating, akgun2021machine, kang2017reinforcement, kang2019q, wei2023reinforcement, li2024page, liaw2024reinforcement, wu2019maximizing, ha2024rl}) apply RL to storage systems, focusing on cloud utilization~\cite{liu2019learning, wang2020reinforcement}, SSD throughput and QoS optimization~\cite{li2023rlalloc}, and maintenance tasks such as garbage collection and error mitigation~\cite{yoo2020reinforcement,kang2017reinforcement,kang2018dynamic,li2020mitigating}. \namePaper{} is orthogonal to these works and can be integrated with them in HSS.

\head{RL for Architectural Optimizations}
RL has also been widely applied to architectural optimization, including data prefetching (e.g., \cite{bera2021pythia,peled2015semantic, bera2026athena}), \camii{memory scheduling (e.g., \cite{ipek2008self, multi_scheduler_HPCA_2012, martinezISCA50Retrospective})}, cache replacement strategies~\cite{liu2020imitation}, and network-on-chip arbitration~\cite{rl_NOC_AIDArc_2018,lin2020deepNOC}. 
These works demonstrate that RL effectively learns control policies in complex hardware-managed environments. \namePaper{} extends this line of research to optimize cross-device data management in HSS. 

\head{Multi-Agent RL Systems}
Several prior works (e.g.,~\cite{jain2017coordinated, qiu2022reinforcement, qiu2022simppo, qiu2020firm, mao2023multi, mlsys2023qiu, lu2022rlrp, jain2017cooperative, shen2022multiagent, sun2025fleetio}) use multi-agent RL (MARL) for system optimization. 
Qiu~\etal~\cite{qiu2022reinforcement, qiu2022simppo, qiu2020firm, mlsys2023qiu} optimize latency in serverless computing platforms.
Jain~\etal~\cite{jain2017coordinated, jain2017cooperative} optimize CPU cache management policies.
Shen~\etal~\cite{shen2022multiagent} address cache cleaning in \camii{shingled magnetic recording (SMR)} drives.
\camii{These works use multiple agents to scale decision-making across similar resources (e.g., compute nodes, cache sets, storage regions in SMR drives). 
Because each agent solves a similar subproblem with a shared objective and operates at the same timescale as other agents, these works use identical agents with shared action spaces and reward functions.
In contrast, HSS data placement and migration differ in their objectives and execution timescale (see \S\ref{subsec:mechanism_challenges}). 
Therefore, \namePaper{} uses heterogeneous RL agents with different action and reward structures. 
}

RLRP~\cite{lu2022rlrp} proposes a deep RL-based replica placement and migration technique for distributed storage systems.
\camii{RLRP (1) operates at coarse-grained time intervals (e.g., only when a new device is added to the system topology), (2) retrains each RL agent \emph{only} when the system topology changes (e.g., when a new device is added or a device fails), and (3) only uses device-related state features (e.g., utilization of storage devices in the network).
In contrast, \namePaper{} performs fine-grained page placement and migration in an HSS while continuously updating its policies online to adapt to rapidly changing workload conditions.
\namePaper{} demonstrates that MARL formulation with heterogeneous agents is beneficial when different tasks (e.g., placement and migration) operate at different timescales and objectives, complementing prior MARL systems that primarily replicate identical agents across similar resources.}

\section{Conclusion}
We introduce \namePaper{}, the first multi-agent reinforcement learning based holistic data-management technique that jointly optimizes data placement and \camii{data} migration \camii{in hybrid storage systems}. \namePaper{} uses two coordinated RL agents that operate on a shared state representation, but optimize complementary objectives at different timescales. 
The data-placement agent selects the best-fit storage device to place I/O request data on the critical path, while the data-migration agent proactively reorganizes previously placed data during low-utilization periods to improve long-term system performance. 
Through coordinated reward structures and lightweight state representations, \namePaper{} enables effective cooperation between the two agents \camii{at low latency and storage overheads}.  
Our real-system evaluation demonstrates that \namePaper{} significantly outperforms \camii{eight} prior \camii{and new} HSS data-management techniques on four HSS configurations across a wide range of workloads. \namePaper{} is easily extensible to different HSS configurations. 
\namePaper{}'s performance benefits come with low latency and storage overheads.
\camii{We hope and believe \namePaper{} will inspire future
work and ideas on self-optimizing storage and memory systems that jointly optimize multiple tasks across heterogeneous devices and workload conditions.}

\begin{acks}
We thank the anonymous reviewers of ISCA 2024, MICRO 2024, HPCA 2025, ISCA 2025, MICRO 2025, FAST 2026, and ICS 2026 for their feedback. 
We thank the SAFARI Research Group members for providing a stimulating \camii{and} inclusive, intellectual, and scientific environment \camii{that enables the flourishing of many novel ideas}.
We acknowledge the generous gifts from our industrial partners, including Google, Huawei, Intel, Microsoft, and VMware.
This research was partially supported by \camii{the} Semiconductor Research Corporation (SRC), ETH Future Computing Laboratory (EFCL), Huawei ZRC Storage Team, and AI Chip Center for Emerging Smart Systems Limited (ACCESS). 
Jisung Park was supported by the NRF (RS-2025-00519994, RS-2023-00283799).
\end{acks}

\bibliographystyle{unsrt}

\end{document}